\documentclass[fleqn, usenatbib,useAMS]{mnras}

\usepackage{graphicx}
\usepackage{amsmath,amssymb}
\usepackage{color}
\usepackage{hyperref}
\usepackage{multicol}
\usepackage{multirow}
\usepackage{mathtools}
\usepackage{siunitx}
\usepackage[dvipsnames]{xcolor}
\usepackage[left]{lineno}
\usepackage{pdflscape}	
\usepackage{subfigure}
\usepackage{multicol}
\usepackage{cuted}
\usepackage{widetext}

\usepackage[T1]{fontenc}
\usepackage{ae,aecompl}
\usepackage{newtxtext,newtxmath}
\usepackage{lipsum}

\newcommand{\beq}{\begin{equation}}
\newcommand{\eeq}{\end{equation}}
\newcommand{\bdm}{\begin{displaymath}}
\newcommand{\edm}{\end{displaymath}}

\definecolor{Gray}{gray}{0.9}

\newcommand{\alm}{a_{\ell, m}}
\newcommand{\blm}{b_{\ell, m}}
\newcommand{\ylm}{Y_{\ell, m}}
\newcommand{\cg}{Clebsch–Gordan }
\newcommand{\sinc}{\text{sinc} }
\newcommand{\gw}{\text{GW}}
\newcommand{\redtext}[1]{{#1}}

\graphicspath{{./figures/}}

\title[Mapping the sky with LISA]{Mapping the Gravitational-wave Sky with LISA: A Bayesian Spherical Harmonic Approach}

\author[Banagiri et al.]{Sharan Banagiri$^{1}$\thanks{E-mail: banag002@umn.edu}, Alexander Criswell$^{1}$, Tommy Kuan$^{1}$, Vuk Mandic$^{1}$, \newauthor Joseph D. Romano$^{2}$, Stephen R. Taylor$^{3}$\\
$^{1}$ School of Physics and Astronomy, University of Minnesota, Minneapolis, Minnesota 55455, USA\\
$^{2}$ Texas Tech University, Physics \& Astronomy Department, Box 41051, Lubbock, Texas 79409, USA \\
$^{3}$ Department of Physics \& Astronomy, Vanderbilt University, 2301 Vanderbilt Place, Nashville, TN 37235, USA \\}

\pubyear{2021}

\begin{document}
\label{firstpage}
\pagerange{\pageref{firstpage}--\pageref{lastpage}}
\maketitle

\numberwithin{equation}{section}

\begin{abstract}
The millihertz gravitational-wave frequency band is expected to contain a rich symphony of signals with sources ranging from galactic white dwarf binaries to extreme mass ratio inspirals. Many of these gravitational-wave signals will not be individually resolvable. Instead, they will incoherently add to produce stochastic gravitational-wave confusion noise whose frequency content will be governed by the dynamics of the sources. The angular structure of the power of the confusion noise will be modulated by the distribution of the sources across the sky. Measurement of this structure can yield important information about the distribution of sources on galactic and extra-galactic scales, their astrophysics and their evolution over cosmic timescales. Moreover, since the confusion noise is part of the noise budget of LISA, mapping it will also be essential for studying resolvable signals. In this paper, we present a Bayesian algorithm to probe the angular distribution of the stochastic gravitational-wave confusion noise with LISA using a spherical harmonic basis. We develop a technique based on \cg coefficients to mathematically constrain the spherical harmonics to yield a non-negative distribution, making them optimal for expanding the gravitational-wave power and amenable to Bayesian inference. We demonstrate these techniques using a series of simulations and analyses, including recovery of simulated distributed and localized sources of gravitational-wave power. We also apply this method to map the gravitational-wave foreground from galactic white-dwarfs using a simplified model of the galactic white dwarf distribution.

\end{abstract}

\begin{keywords}
gravitational waves, methods:statistical, methods: data analysis
\end{keywords}

\section{Introduction}
\label{Sec:Intro}

The upcoming space-based Laser Interferometer Space Antenna (LISA)~\citep{amaroseoane2017laser} promises access to the millihertz gravitational-wave (GW) frequency band, which is inaccessible to terrestrial ground-based detectors like the Laser Interferometer Gravitational-wave Observatory (LIGO). A rich collection of galactic and extra-galactic sources emit GWs at these frequencies. Among them are double white dwarfs (DWDs) from both within the Milky Way~\citep{Marsh_2011, Korol:2017qcx} and neighboring satellite galaxies~\citep{Korol:2018ulo, Korol:2020lpq}, extreme mass ratio inspirals~\citep{AmaroSeoane:2007aw, Gair:2017ynp, Babak:2017tow}, supermassive blackhole binaries~\citep{Barausse:2014oca, Klein:2015hvg}, extragalactic stellar-mass binary black hole (BBH) and binary neutron star inspirals~\citep{Lau:2019wzw}, and even exoplanets orbiting white dwarfs~\citep{Tamanini_Danielski_2019, Danielski:2019rvt}. Cosmological sources like cosmic strings~\citep{Auclair:2019wcv}, cosmological phase transitions~\citep{Caprini:2015zlo} and primordial GW backgrounds~\citep{Bartolo:2016ami} from the early universe are also potentially accessible by LISA, not to mention the prospect of multi-wavelength GW science in conjunction with next-generation ground-based detectors (see for example~\cite{Axen:2018zvb, Lasky:2015lej})

The GWs from many astrophysical sources will not be individually detectable and will overlap to form a confusion noise in the detector, usually called the stochastic gravitational-wave background~\citep{Regimbau:2011rp, Romano:2016dpx}. The stochastic GW background from binary inspirals is a major scientific target for both current ground-based detectors like Advanced LIGO~\citep{TheLIGOScientific:2014jea} and Advanced Virgo~\citep{TheVirgo:2014hva}, and for pulsar timing arrays~\citep{Brazier:2019mmu, Manchester:2012za, Kramer:2013kea}. The current limits from ground-based detectors are \redtext{$\Omega_{\gw} (f = 25 \text{ Hz}) < 3.4 \times 10^{-8}$ ~\citep{Abbott:2021xxi} for an isotropic background from compact binary coalescence } and \redtext{$h_0 < (1.7 - 2.1) \times 10^{-25}$ for point sources~\citep{Abbott:2021jel}}. The latest results from pulsar timing arrays are from the NANOGrav collaboration which see strong evidence for a common-spectrum red-noise process with a median strain of $1.92 \times 10^{-15}$ at a frequency of $1 \text{ yr}^{-1}$~\citep{Arzoumanian:2020vkk}.

\redtext{There has been a considerable theoretical effort in recent years to model the angular power spectrum of the stochastic background from astrophysical sources, primarily in the sensitive frequency band of ground-based detectors~\citep{Cusin:2018rsq, Jenkins:2018uac, Jenkins:2019nks, Alonso:2020rar, Bartolo:2019zvb}, but also in the millihertz LISA band~\citep{Cusin:2019jhg}. These studies suggest that the degree of anisotropy in the extragalactic astrophysical stochastic background is relatively small compared to the monopole. On the other hand, the anisotropy in the stochastic background from supermassive black hole binaries in the pulsar timing array band is expected to be much larger~\citep{Mingarelli:2017fbe}.} 
 
 In the case of LISA, there is also a GW foreground from galactic DWDs that stands above the instrumental noise for a part of the LISA band while still being stochastic in nature. While this foreground is considered an inconvenient noise source for resolvable signals, it is of astrophysical interest in its own right and contains useful information about the physical and spectral distribution of the DWDs~\citep{Benacquista:2005tm, Breivik:2019oar}. For the rest of the paper, we will use the term stochastic gravitational-wave confusion noise (SGCN) to refer to both stochastic backgrounds and foregrounds collectively.
 
 The angular structure of the GW power from a SGCN will directly follow the distribution of the sources which generates it. The antenna patterns of the detectors are not isotropic and also change throughout LISA's orbit, which ensures that the angular structure can, in principle, be measured with enough integration time. Spherical harmonic functions are a natural basis to describe the distribution of power on the sky and have been frequently used in algorithms developed to measure these anisotropies, both for LIGO and LISA, albeit in a frequentist manner~\citep{Thrane:2009aa, Ungarelli:2001xu, Kudoh:2004he, Taruya:2005yf, Taruya:2006kqa, Renzini:2018vkx}. In particular, recently \cite{Contaldi:2020rht} developed a frequentist maximum likelihood method to map GW power with LISA. 
 
In this paper, we present a Bayesian algorithm to map the power of an SGCN using a spherical harmonic basis. There are several advantages to developing a Bayesian version of this method, especially in the case of LISA, where the galactic foreground dominates. First, the Bayesian version can be better integrated with global analyses designed to extract multiple resolvable signals~\citep{Littenberg:2020bxy}, in order to map the foreground simultaneously along with them. Accounting for the foreground in this way can be crucial for accurately inferring the properties of the resolvable signals. Frequentist searches also generally require the inversion of a Fisher matrix connecting different sky directions or harmonics. The poor angular sensitivity of GW detectors, LISA included, creates degeneracies that make this inversion mathematically ill-conditioned, necessitating the use of techniques like singular value decomposition (see for example~\cite{Thrane:2009aa, Contaldi:2020rht}). These degeneracies are better accommodated with a Bayesian approach, which requires no such inversion. Additionally, the angular sensitivity of the spherical harmonic expansion is set by cutting off the expansion in $\ell$ at some $\ell_{\text{max}}$ parameter value. This parameter value is chosen in a somewhat ad hoc way in frequentist searches but can be much more naturally accommodated in Bayesian searches by allowing the data to determine it. 

Historically, one hindrance of a Bayesian spherical harmonic implementation has been that the generic expansion describes a complex field on the sky, while GW power is non-negative by definition. We demonstrate a way to mathematically impose this constraint using \cg coefficients in a Bayesian spherical harmonic analysis. This mathematical technique was used recently to measure the angular distribution of GW detections by LIGO-Virgo~\citep{Payne:2020pmc}, while a similar method was also recently used in the pulsar timing array band~\citep{Taylor:2020zpk}. We also introduce the Bayesian LISA Pipeline (BLIP) designed to simulate LISA data, perform the spherical harmonic analysis on the simulated data, and conduct Bayesian inference to recover the simulated parameters. 
 
 The rest of the paper is structured as follows. In Sec.\ref{Sec:GW_power} we review SGCNs and the spherical harmonic basis and calculate the detector response function of LISA to GW power in the spherical harmonic basis. In Sec.~\ref{Sec:CG_decomp} we calculate the \cg decomposition for non-negative fields and develop the parameterization necessary for Bayesian inference. Sec.~\ref{Sec:Blip} introduces the BLIP pipeline and discusses the likelihood function and the configuration used for the analyses in this paper. \redtext{Sec.~\ref{Sec:Sim_Det} demonstrates measurement of anisotropies in simulated LISA data in the spherical harmonic basis with the \cg decomposition. Sec.~\ref{Sec:Foreground} discusses applications of the technique to the galactic foreground followed by a discussion and conclusion in Sec.~\ref{Sec:Conclusion}}.

\section{Stochastic gravitational-wave confusion noise}
\label{Sec:GW_power}

Astrophysical SGCNs result from an incoherent superposition of GWs from many disparate sources which are not individually resolvable~\citep{Regimbau:2011rp}~\footnote{This implies that the strength of the background depends on the sensitivity of the detector. \redtext{While more sensitive detectors are also better at detecting SGCNs, they can also resolve more sources potentially reducing the confusion noise from astrophysical sources. }}. Appealing to the central limit theorem, one can characterize the SGCN as colored Gaussian noise in the detectors~\citep{Allen:1997ad}. This is usually a good assumption for sources in the LISA band that overlap with one other. The metric perturbation at $(t, \textbf{x})$ corresponding to an SGCN can be written in the Fourier basis as, 

\begin{equation}
\begin{split}
	h_{ij} (t, \textbf{x}) = \sum_A \int^{\infty}_{-\infty} df & \int  d^2 n \, \tilde{h}_A(f, \textbf{n}) \, e^A_{ij} (\textbf{n}) \\ & \times \,\exp \{ -2 \pi i f  (t - \textbf{n} \cdot \textbf{x}/c) \} , 
\end{split}
\label{Eq:metric_perturb}
\end{equation}
where $A = \{ + , \times \}$ denotes polarization, $c$ is the speed of light, $\textbf{n}$ is the directional unit vector, $e^A_{ij} (\textbf{n}) $ are the polarization tensors \redtext{ and $i, j$ denote spatial indices}. The frequency of the SGCN is represented by $f$ and $\tilde{h}_A(f, \textbf{n})$ are the Fourier components of the perturbations which satisfy the following, 

\begin{equation}
\begin{split}
		\langle \tilde{h}_A(f, \textbf{n}) \rangle = & \, 0,  \\ \langle \tilde{h}_A(f, \textbf{n}) \, \tilde{h}^*_{A'}(f', \textbf{n}') \rangle = & \frac{1}{2}\, S_A (f, \textbf{n}) \, \delta_{A, A'} \, \delta(f - f')\, \delta^2( \textbf{n}, \textbf{n}'). 
\end{split}
\end{equation}

Under the assumption of Gaussianity the power spectrum $ S_A (f, \textbf{n})$ is the main measurable quantity of an SGCN. For the rest of this paper we will also assume that the SGCN is unpolarized i.e, $S_+(f, \textbf{n}) = S_{\times} (f, \textbf{n}) = 1/2 \, S_{\gw} (f, \textbf{n})$. The power spectrum is conventionally characterized by the dimensionless energy density $\Omega_{\gw} (f, \textbf{n})$ per logarithmic frequency bin~\citep{Allen:1997ad}, related to $S_{\gw} (f, \textbf{n}) $ via,

\begin{equation}
	\Omega_{\gw} (f, \textbf{n}) = \frac{2 \pi^2 f^3}{3 H^2_0} S_{\gw} (f, \textbf{n}), 
	\label{Eq:GW_power}
\end{equation} 
where $H_0$ is the Hubble constant. In general, the distribution of the power on the sky will not be isotropic but will rather trace the distribution of its sources. To describe the angular structure of the SGCN we will first assume that its frequency and directional dependence can be factorized as

\begin{equation}
	\Omega_{\gw} (f, \textbf{n}) = \Omega (f) \mathcal{P}(\textbf{n}). 
	\label{Eq:factorization}
\end{equation}
 The spectral shape of the SGCN is given by $\Omega (f)$ while $\mathcal{P}(\textbf{n})$ describes the angular distribution of the background, normalized so that: 

\begin{equation}
	\int_{\mathcal{S}^2} d^2 n \, \mathcal{P}(\textbf{n}) = 1.
	\label{Eq:norm}
\end{equation}

A common way to parameterize the spectral shape $\Omega (f)$ is to write it as a power law:

\begin{equation}
	\Omega (f) = \Omega_{\text{ref}} \left ( \frac{f}{f_\text{ref}} \right )^{\alpha}, 
	\label{Eq:Power_law}
\end{equation}
where $\alpha$ is the spectral index of the power law, $f_\text{ref}$ is some reference frequency and $\Omega_{\text{ref}} = \Omega (f=f_\text{ref}) $. In particular stochastic backgrounds and foregrounds from compact binaries are expected to follow a power law with $\alpha = 2/3$~\citep{Phinney:2001di}. For the rest of this paper, we will assume the power-law spectral shape to hold. 

Spherical harmonics provide a general orthonormal basis to parameterize an arbitrary continuous and differentiable function on the two sphere:

\begin{equation}
    \mathcal{P} (\textbf{n})  = \frac{1}{\sqrt{4 \pi a_{0, 0}}} \sum_{\ell, m} \alm \ylm (\textbf{n}), 
    \label{Eq:Sph_prior}
\end{equation}
where $\{\ylm \}$ are the spherical harmonic functions and the normalization factor of $\sqrt{4\pi a_{0, 0}}$ ensures that the expansion satisfies Eq.~\ref{Eq:norm}. The coefficients $\{\alm\}$ are in general complex numbers and characterize the distribution of the field on the sky. The harmonics of positive and negative $m$ are related by:

\begin{equation}
 Y_{\ell, - m} (\textbf{n})  = (-1)^m \ylm^* (\textbf{n}), 
\end{equation}
The spherical harmonic functions form an orthonormal basis on the two sphere, 

\begin{equation}
	\int d^2 n \, \ylm (\textbf{n}) Y^*_{\ell', m'} (\textbf{n}) = \delta_{\ell, \ell'} \delta_{m, m'}
\end{equation}

\subsection{Detector response in the spherical harmonic basis}
\label{Sec:Det_response}

\redtext{The currently proposed LISA space mission consists of three satellites orbiting around the Solar System barycenter in an approximately triangular formation~\citep{amaroseoane2017laser}. The sides of the triangle will be about $2.5 \times 10^6$ \text{km} with the centroid following the same orbit as the Earth. Three Michelson channels can be formed at each vertex by considering the differential strain between the two arms containing the vertex. Since any two channels will share an arm with each other, the noise in the three channels is not independent of each other.}

\redtext{The default Michelson channels suffer from laser phase noise which can swamp out any possible GW signal. By time-shifting and linearly combining the Michelson channels, one can generate data combinations where the laser phase noise is canceled out, a process called time-delay interferometry (TDI).  Two generations of such TDI combinations are generally used. The first generation TDI channels are usually called X, Y, and Z and work best when the arm lengths of the LISA constellation can be approximated to be constant. The second-generation TDI channels, sometimes called A, E, and T, have also been developed, which work as well for slowly evolving arm lengths. We point to~\cite{Tinto:2004wu} for a detailed overview of TDI.}

\redtext{The strain data $d_I (t)$ from any channel $I$ is the sum of the instrumental noise $n_{I} (t)$ and the GW signal in the channel $h^{GW}_{I} (t)$,}

\begin{equation}
	d_I (t) = n_I (t) + h^{GW}_I (t).
\end{equation}
\redtext{Due to the linear nature of the Fourier transform this relationship carries over to the frequency domain as well,} 
\begin{equation}
	\tilde{d}_I (f) = \tilde{n}_I (f) + \tilde{h}^{GW}_I (f).
\end{equation}
\redtext{In the case of SGCNs, due to the Gaussian nature of both the GW strain and the instrumental noise and under the assumption that they are uncorrelated, the PSD of the data is the sum of the signal and the noise PSDs, }

\begin{equation}
	S_{II}(f) = S^n_{II} (f) + S^{GW}_{II} (f). 
	\label{Eq:Autocorr}
\end{equation} 
\redtext{A similar relation will also hold for cross correlation between channels $I$ and $J$,} 
\begin{equation}
	S_{IJ}(f) = S^n_{IJ} (f) + S^{GW}_{IJ} (f). 
	\label{Eq:Crosscorr}
\end{equation}

Since we decompose GW power in the orthonormal spherical harmonic basis, it is also useful to calculate the detector response to each spherical harmonic mode. First, we define the antenna pattern function of channel $I$ of the detector as~\citep{Cornish:2001qi,  Cornish:2001bb},

\begin{equation}
	F^A_I (f, t, \textbf{n}) = D_I (f, t, \textbf{n}) : e^A (\textbf{n}),
\end{equation}
where $D_I (f, t, \textbf{n})$ is the detector response tensor. 

For a Michelson channel \redtext{with vertex at $\textbf{r}$ and} arm orientations given by unit vectors $\textbf{u}$ and $\textbf{v}$, the response tensor is given by, 

\begin{equation}
\begin{split}
		D_{\text{mich}} (f, t, \textbf{n}) = \frac{1}{2} \bigg [ &\left( \textbf{u} \, \otimes \, \textbf{u} \right) \mathcal{T} (f, \textbf{u} \cdot  \textbf{n}) \\ & - \left( \textbf{v} \, \otimes \, \textbf{v} \right) \mathcal{T} (f, \textbf{v} \cdot  \textbf{n}) \bigg ] \exp \left \{ 2 \pi i f \textbf{n} \cdot \textbf{r}/c \right \},
\end{split}
\end{equation}
assuming that the satellite motion is negligible during the round-trip light-travel time between the satellites. The response tensor is a function of time owing to the temporal variation of $\textbf{u}$ and $\textbf{v}$ as the satellites move in their orbits. Here $\mathcal{T} (f, \textbf{u} \cdot  \textbf{n})$ is the timing transfer function of interferometric detectors to GWs, which for an equal arm detector is given by~\citep{Schilling_1997, Cornish:2001qi}:

\begin{equation}
\begin{split}
\mathcal{T} (f, \textbf{u} \cdot  \textbf{n}) & = \frac{1}{2} \bigg [ \sinc\left( \frac{f}{2 f_*} (1 - \textbf{n} \cdot \textbf{u})\right)  \exp \left (- i \frac{f}{2 f_*}  ( 3 + \textbf{n} \cdot \textbf{u} ) \right)  \\	
			+ & \, \sinc\left( \frac{f}{2 f_*}  (1 + \textbf{n} \cdot \textbf{u})\right)  \exp \left ( - i  \frac{f}{2 f_*}  ( 1 + \textbf{n} \cdot \textbf{u} ) \right) 	\bigg ],
\end{split}
\end{equation}
where $L = 2.5 \times 10^9$ m is the arm length of LISA and {$f_* \equiv c/ 2 \pi L $}. Since the SGCN is measured as the excess GW power in a detector, to detect it, we will need to correlate two channels $I$ and $J$~\footnote{Note that $I$ and $J$ can be the same channel of LISA. In the case of LIGO and Virgo, we usually only correlate distinct spatially-seperated interferometers, so the correlation response function is called the overlap reduction function}. \redtext{Assuming LISA's motion can be neglected, the GW strain signal in channel $I$ from the metric perturbation defined in Eq.~\ref{Eq:metric_perturb} will be}

\begin{equation}
	h_{I} (t, \textbf{x}) = \sum_A \int^{\infty}_{-\infty} df  \int  d^2 n \, \tilde{h}_A(f, \textbf{n}) \, F^A_{I} (f, \textbf{n}) \, e^{ -2 \pi i f t } .
\end{equation}

\redtext{Cross-correlating data between two channels I and J leads to the response function to the GW power distribution $\mathcal{P}(\textbf{n})$  on the sky,}

\begin{equation}
	R^{IJ}(f, t) = \int \frac{d^2 n}{4 \pi} \, \mathcal{P} (\textbf{n}) \left( \sum_A F^A_I (f, \textbf{n})  {F^A_J}^* (f, \textbf{n}) \right).
\end{equation}
Expanding $\mathcal{P}(\textbf{n})$ in the spherical harmonic basis as in Eq.~\ref{Eq:Sph_prior} we can define the response function to the $\ylm$ mode to be,

\begin{equation}
	\mathcal{R}_{\ell, m}^{IJ} (f, t) = \int \frac{d^2 n}{4 \pi} \,  Y_{\ell, m} (\textbf{n}) \left( \sum_A F^A_I (f, \textbf{n})  {F^A_J}^* (f, \textbf{n}) \right).
		\label{Eq:Sph_response}
\end{equation}

Finally combining Eqs.~\ref{Eq:GW_power}, \ref{Eq:factorization}, \ref{Eq:Sph_prior} and \ref{Eq:Sph_response}, the SGCN GW power in the correlation between channels $I$ and $J$ is,

\begin{equation}
	S_{IJ}^{\text{GW}} (f, t) = \frac{3 H^2_0}{2 \pi^2 f^3} \frac{\Omega (f) }{\sqrt{4 \pi a_{0, 0}}} \sum_{\ell, m} \alm \mathcal{R}_{\ell, m}^{IJ} (f, t) .
	\label{Eq:PSD_sgwb}
\end{equation}

\section{Clebsch-Gordan decomposition}
\label{Sec:CG_decomp}

The general spherical harmonic expansion describes a complex field. We can constrain it to be real everywhere on the sky with the condition:

\begin{equation}
	a_{\ell, -m} = (-1)^m \alm^* . 
\end{equation}
However, the decomposition of GW power needs to be not only real but also non-negative for any direction on the sky, i.e., $	\Omega(f, \textbf{n}) \geq 0$. This is especially important for LISA, as the GW power distribution will be highly anisotropic due to the foreground from galactic binaries. Implementing this constraint is also necessary for Bayesian inference with the spherical harmonic basis. This is because the posterior should be zero for any set of $\alm$'s that contains even the tiniest spot on the sky with negative GW power. Previous work in the PTA band attempted to solve this problem by numerically checking the sign of the GW power on the sky using a grid~\citep{Taylor:2013esa, Taylor:2015udp} and assigning a probability of zero for a given set of $\alm$'s if any of the pixels have negative power. However, one can always use finer and finer grids  to check this which makes this solution computationally untenable. A solution that mathematically guarantees non-negative power is much preferable. In this section, we describe a solution to this problem using \cg coefficients. 

First we define a function $\mathcal{S} (\textbf{n})$ which is the square root of the spherical harmonic expansion, i.e,

\begin{equation}
	\mathcal{S}(\textbf{n}) = \left [  \sum_{\ell, m} \alm \ylm(\textbf{n}) \right ]^{1/2}.
	\label{Eq:S_define}
\end{equation}\\
We then expand $\mathcal{S} (\textbf{n})$ through its own spherical harmonic expansion as,

\begin{equation}
	    \mathcal{S}(\textbf{n}) = \sum_{\ell, m} \blm \ylm(\textbf{n}). 	 
		\label{Eq:S_expansion}
\end{equation}
The necessary and sufficient condition for the GW power to be non-negative then becomes $\mathcal{S}(\textbf{n}) \in \mathbb {R}$ which implies that $ b_{\ell, - m} = (-1)^m \blm^*$. From Eqs.~\ref{Eq:S_define} and~\ref{Eq:S_expansion} we get,

\begin{equation}
 \sum_{L, M}a_{L, M} Y_{L, M} =   \left(\sum_{\ell, m} \blm \ylm (\textbf{n}) \right)^2 .
\end{equation}
Expanding the right-hand side gives us:

\begin{equation}
\sum_{L, M}a_{L, M} Y_{L, M}  = \sum_{\ell,  m}\sum_{\ell',  m'}  \blm b_{\ell',  m'} \ylm (\textbf{n}) Y_{\ell', m'} (\textbf{n}).
    \label{Eq:alm_blm_exp}
\end{equation} 

\redtext{The  Clebsch-Gordan coefficients  $C^{LM}_{\ell m, \ell' m'}$ enable us to write products of spherical harmonics as a sum over spherical harmonics~\footnote{Alternatively one can use Wigner-3j symbols}, a trick that has been used in GW literature before but in different contexts~\citet{Cornish:2002bh, Cornish:2001hg}}. Here we get ;

\begin{equation}
\begin{split}
 Y_{\ell, m}(\textbf{n}) Y_{\ell', m'} (\textbf{n}) =  \sum_{L  = L_{\text{min}}}^{L_{\text{max}}} & \sqrt{\frac{(2\ell +1)(2\ell' + 1)}{4 \pi (2L + 1)}}  \\ & \times  C^{LM}_{\ell m, \ell' m'} \, C^{L0}_{\ell \, 0, \ell' \, 0}   Y_{L, M} (\textbf{n}).
\end{split}  
\label{Eq:CG_exp}
\end{equation}\\ \\
The expansion obeys selection rules related to the symmetries of the rotation group SO(3). We point to  Ch.~16 of \cite{ArfkenGeorge1972MMfP} for a detailed discussion of the mathematics of spherical harmonics and Clebsch-Gordan coefficients. The selection rules can be listed as: 

\begin{itemize}
    \item  $M = m + m'$
    \item $L_{\text{min}} = \text{min}(|\ell - \ell'|, |m + m'|) $ and $L_{\text{max}} = \ell + \ell'$ 
    \item $L$ is an integer 
\end{itemize}

For compactness, let us define $\beta^{L,M}_{\ell m, \ell' m'}$ such that:

\begin{equation}
    \beta^{L,M}_{\ell m, \ell' m'} = \sqrt{\frac{(2\ell +1)(2\ell' + 1)}{4 \pi (2L + 1)}} C^{LM}_{\ell m, \ell' m'} \, C^{L0}_{\ell \,0, \ell' \, 0}, 
\end{equation}
when the selection rules are satisfied, but $ \beta^{L,M}_{\ell m, \ell' m'} = 0$ otherwise. We can then write Eq.~\ref{Eq:CG_exp} as, 

\begin{equation}
    Y_{\ell, m} (\textbf{n}) Y_{\ell', m'} (\textbf{n}) =  \sum_{L,M} \beta^{L,M}_{\ell m, \ell' m'} Y_{L,M} (\textbf{n}).
\end{equation}
Combining this with equation Eq.~\ref{Eq:alm_blm_exp} gives:

\begin{equation}
\begin{split}
\sum_{L,M}a_{L,M} Y_{L,M} (\textbf{n})  = \sum_{L,M} & \left(\sum_{\ell m}\sum_{\ell' m'}  b_{\ell, m} b_{\ell' , m'}  \beta_{L,M}^{\ell m, \ell' m'} \right) \\ & \times Y_{L,M} (\textbf{n}).
\end{split}
\end{equation}
Since the set of $Y_{LM}$ form an orthonormal basis, this provides the recipe for converting between $\alm$ and $\blm$:

\begin{equation}
    a_{L,M} = \sum_{\ell, m}\sum_{\ell' , m'}  b_{\ell , m} b_{\ell' , m'}  \beta_{L,M}^{\ell m, \ell' m'}.
    \label{Eq:alm_2_blm}
\end{equation}

Often, we want to impose an artificial cutoff on angular sensitivity of some $\ell^a_{\text{max}}$ on the expansion in $\alm$'s. This cutoff can correspond to estimated resolution limits of the detector itself, or it might be astrophysically motivated. The corresponding cutoff of the expansion on the $\blm$'s is taken to be $\ell^b_{\text{max}} = \ell^a_{\text{max}}/2 $ assuming that $\ell^a_{\text{max}}$ is an even number. This is a consequence of the second selection rule that $L_{\text{max}} = \ell + \ell'$, which implies that the cutoff on the expansion in the $\blm$'s should be half that of the $\alm$'s if we want all terms higher than $\ell^a_{\text{max}}$ in the latter to be zero. For the rest of this paper we will use this relation and assume $\ell^a_{\text{max}}$ is even. 

Since $\Omega(f, \textbf{n})$ is proportional to $\mathcal{S}^2(\textbf{n})$, it is invariant under any transformation which leaves the latter invariant. The constraint that $\mathcal{S}(\textbf{n})$ be real~\footnote{In fact we could have chosen $\mathcal{S}$ such that $\Omega = |\mathcal{S}|^2$ in which case $\mathcal{S}$ could be any general complex number. The degree of degeneracies would be higher, so it would be a less economical choice than what we actually make.} makes $\Omega(f, \textbf{n})$ invariant under a parity transformation $\{ \blm \} \rightarrow\{ - \blm \} $. However, this leftover symmetry introduces degeneracies that induce multiple modes in the posterior distribution. Moreover, since the $\{\alm\}$ expansion is normalized as in Eq.~\ref{Eq:norm}, there is also a scale invariance within the $\{\blm\}$ space, i.e with the transformation $\{\blm\} \rightarrow \{\kappa \, \blm\}$, where $\kappa$ is some constant. One can break both these symmetries by fixing the value of one of the $\blm$ coefficients. In this paper we choose to fix $b_{0,0} = 1$. 

A similar method to what is used here can also be applied to expand a probability distribution or any other non-negative function on the two sphere. Most of the results in this section will generalize to that case with the additional requirement that the normalization of the distributions be one. Indeed, the Clebsch-Gordan based spherical harmonic decomposition was recently used to constrain anisotropies in the distribution of BBH progenitors using events detected by LIGO-Virgo up-to-the second observing run in~\cite{Payne:2020pmc}, and also to develop an optimized  anisotropic pipeline for PTAs in \cite{Taylor:2020zpk}.

\begin{figure*}
\subfigure[]
{
    \scalebox{0.5}{\includegraphics{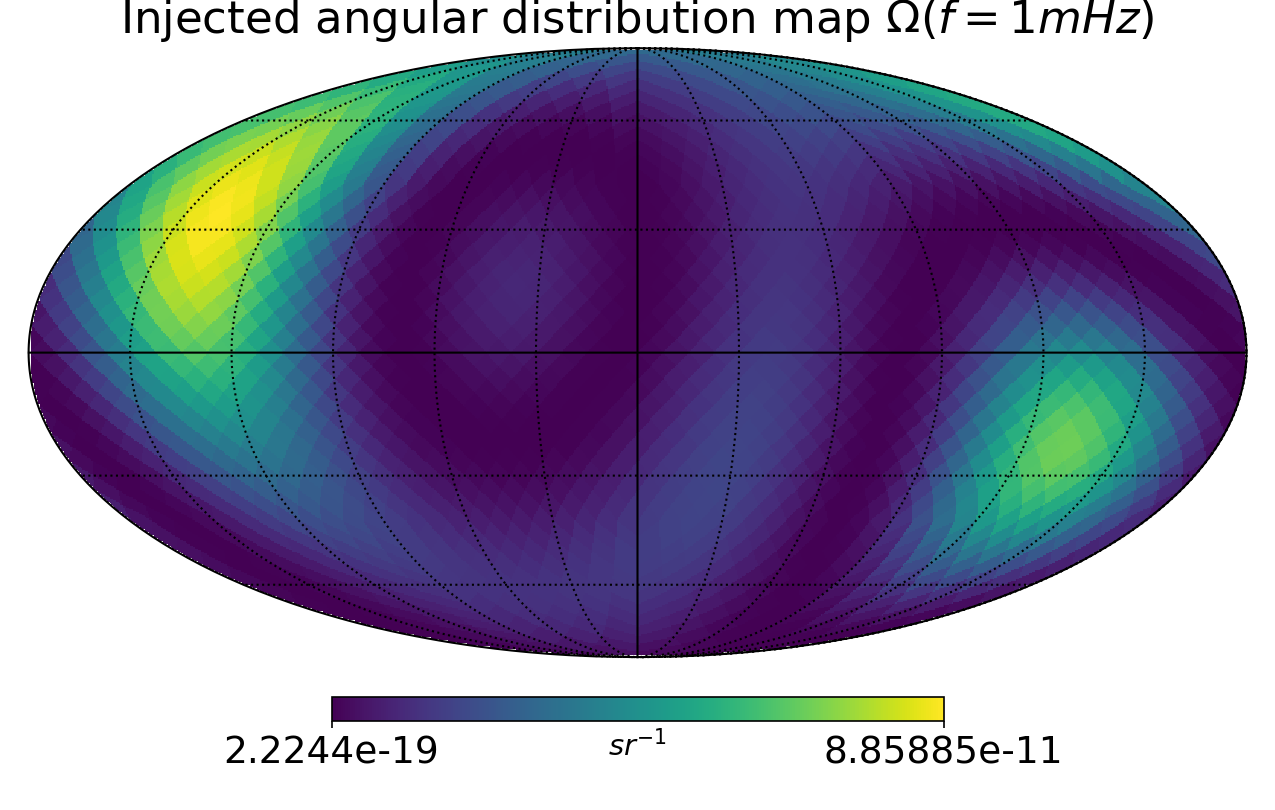}}
}
\subfigure[]
{
    \scalebox{0.5}{\includegraphics{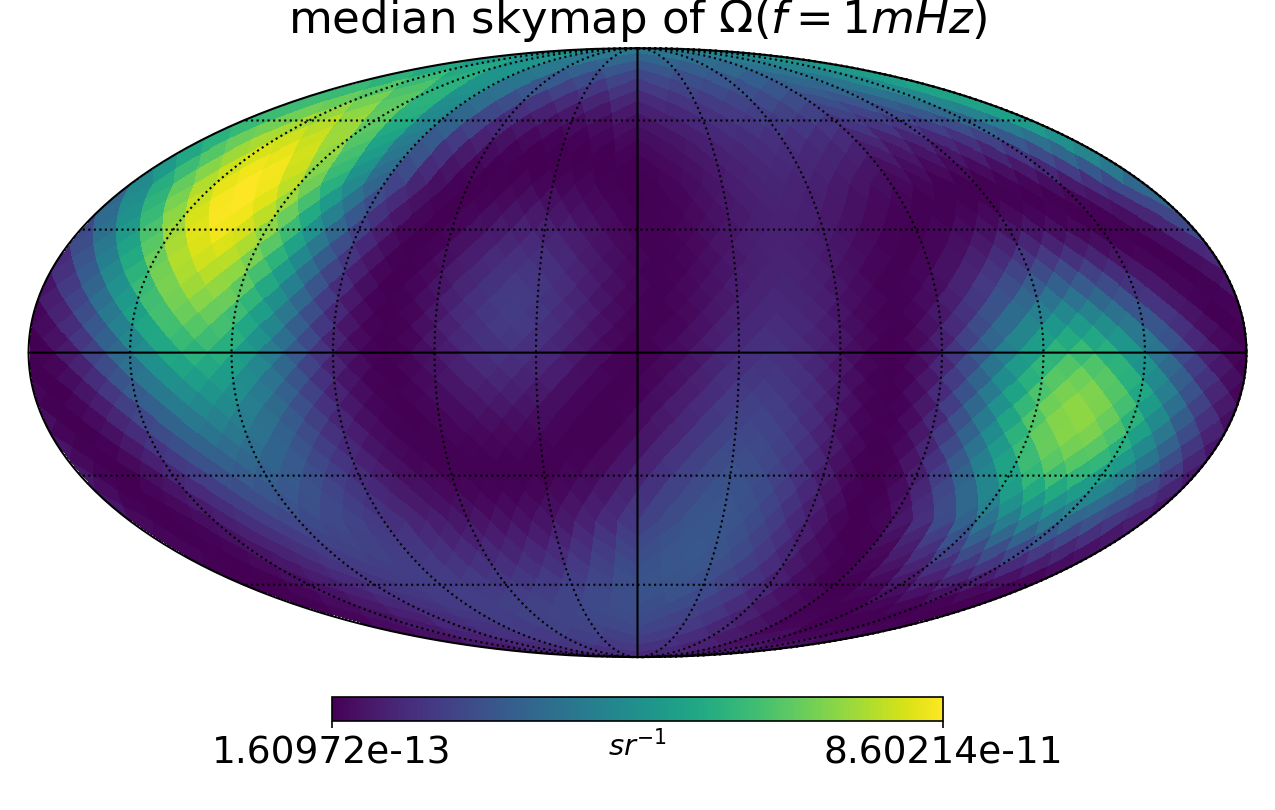}}
}
\subfigure[]
{
    \scalebox{0.5}{\includegraphics{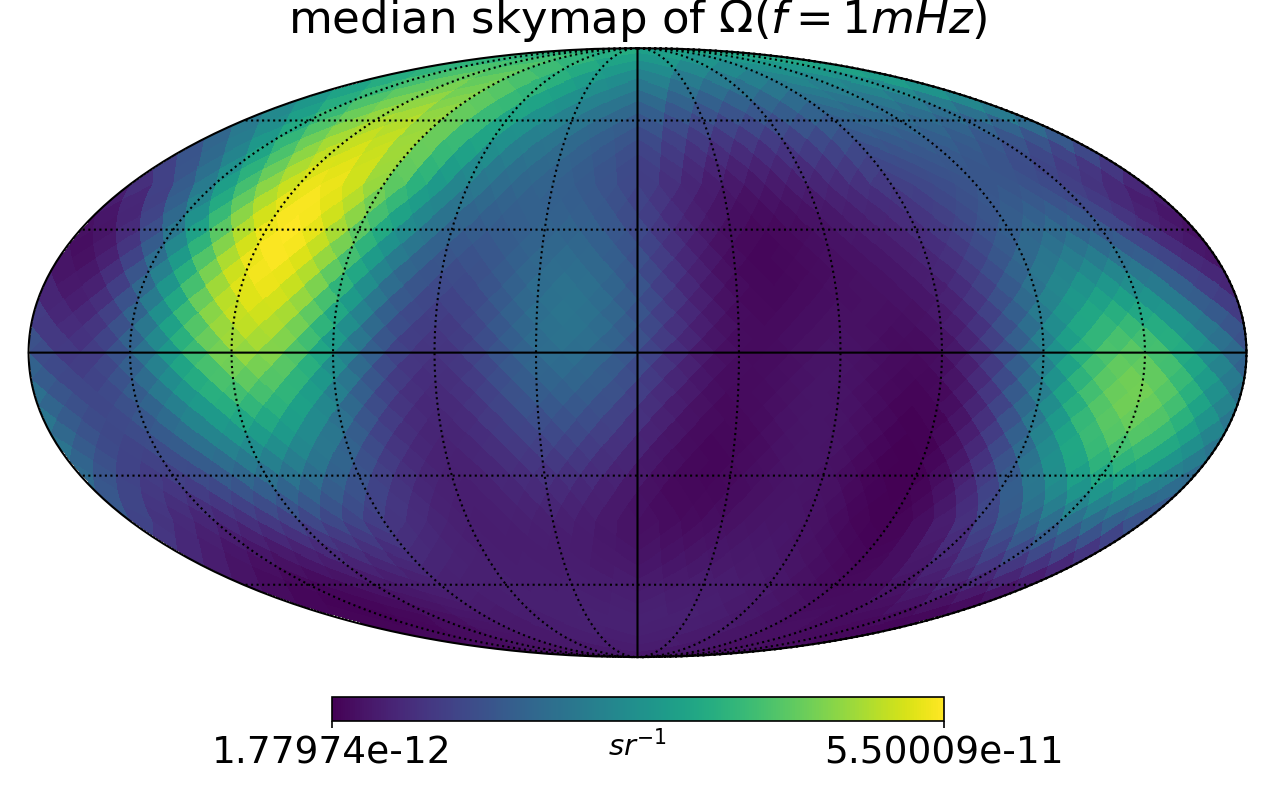}}
}

 \caption{\redtext{Skymaps for the simulation and analysis described in Sec. \ref{Sec:Sim_Det}. Maps b and c show the median posterior sky distribution of $\Omega(f = 1$ mHz) in the solar system barycentric frame, for an analysis time scale of one year and two months, respectively. The injected power was the same in both cases and is shown in map a. The improvement of the one-year recovery compared to the two-month recovery is clear. The full posteriors corresponding to maps b and c are shown in Fig.~\ref{Fig:posteriors_1yr} and Fig.~\ref{Fig:posteriors_2mn} respectively.}}
 \label{Fig:skymaps}
\end{figure*}

\begin{figure*}
\centering
\subfigure[]
{
    \scalebox{0.4}{\includegraphics{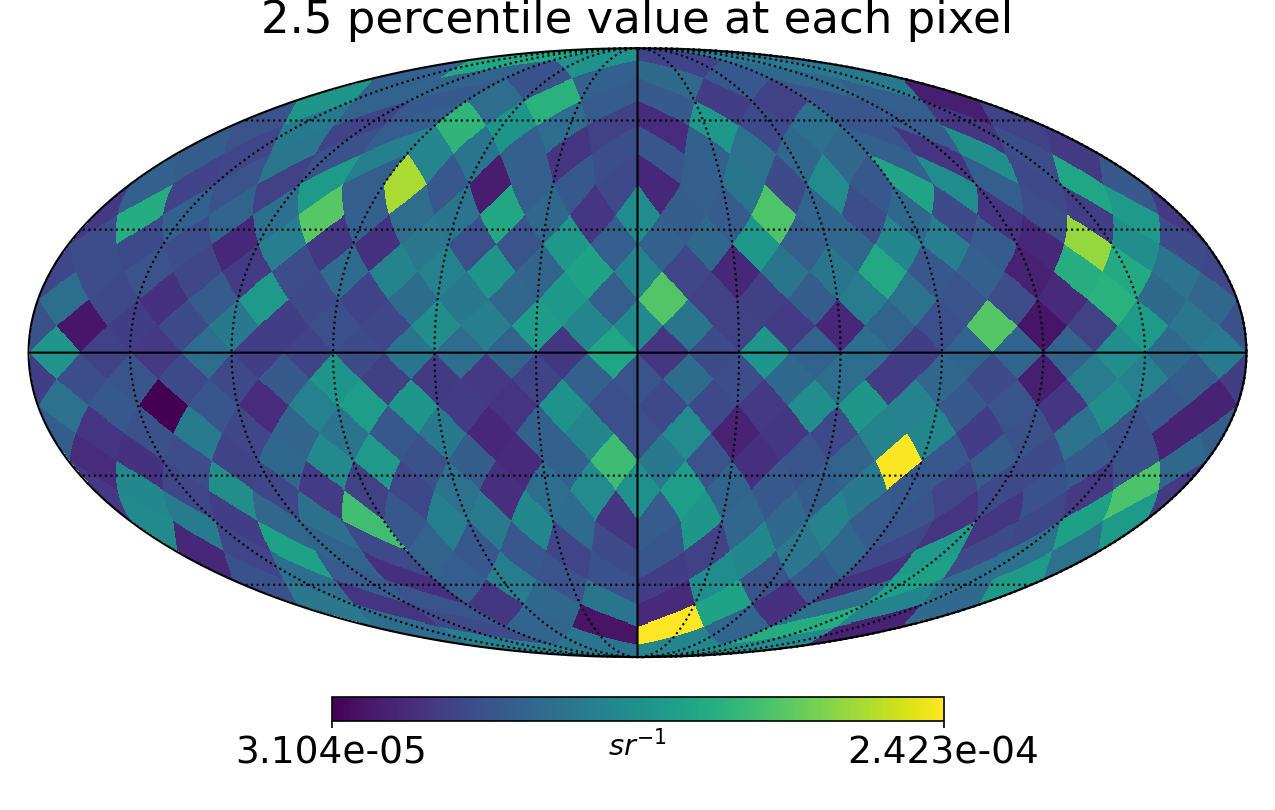}}
}
\subfigure[]
{
    \scalebox{0.4}{\includegraphics{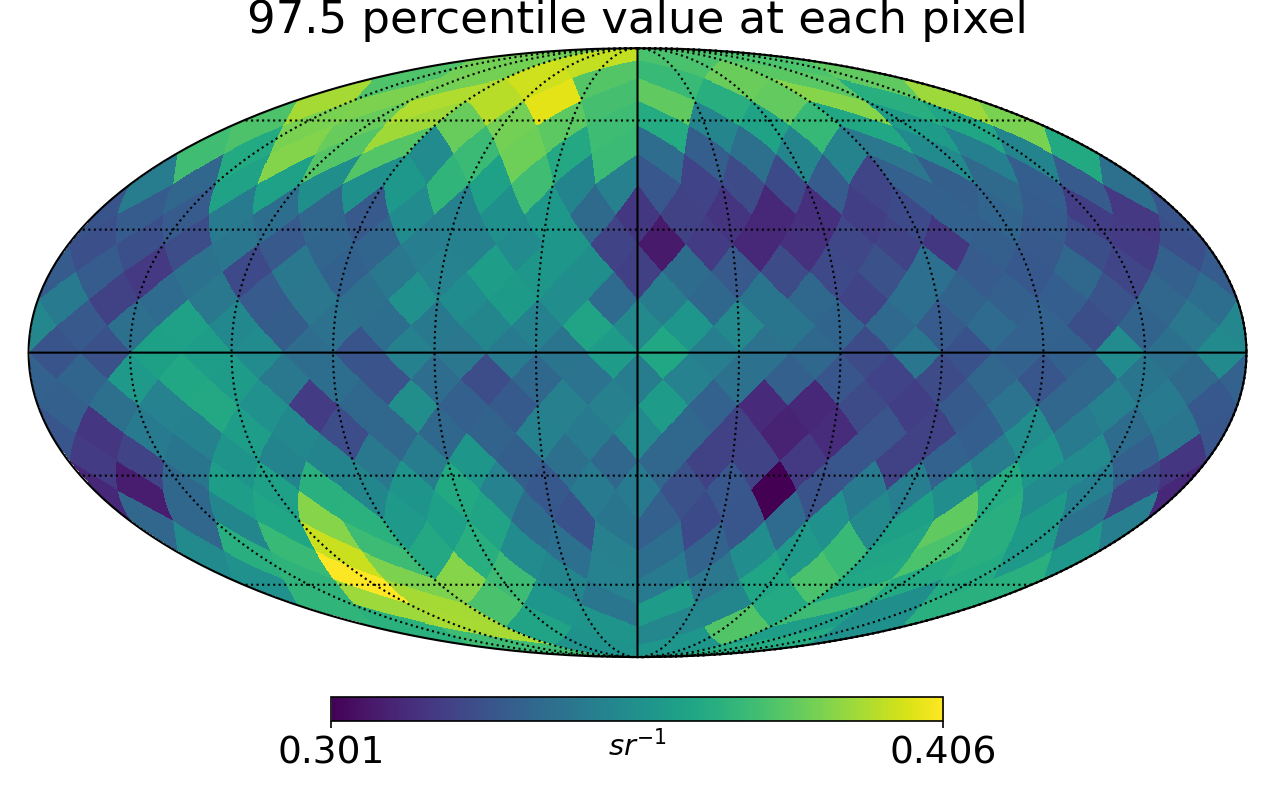}}
}
\caption{\redtext{ The 2.5 and 97.5 percentile values of the prior distribution for each pixel are shown here, without the multiplicative factor of $\Omega (f) $. The range of values show the range of the prior support at each pixel, covering a $95 \%$ confidence interval. For comparison, a uniform distribution on the sky will have a have of value of $2.31 \times 10^{-3} \text{sr}^{-1}$ in each pixel for these HEALPix maps which use $n_{\text{side}} = 6$. The prior thus covers a range of about two orders of magnitude above and below uniform distribution in each pixel. } }	
\label{Fig:prior_skymap}
\end{figure*}

\section{BLIP pipeline}
\label{Sec:Blip}

This section briefly introduces the BLIP pipeline~\footnote{https://github.com/sharanbngr/blip}, which is an independent Python-based implementation for LISA data analysis and the details of it. The pipeline is written to make it easy to add new GW signal models and likelihood models along with simulating instrumental Gaussian noise in the time domain. The instrumental noise is simulated through the acceleration and position noises using the spectral form described in the LISA proposal~\citep{amaroseoane2017laser}.  The functional forms for the power spectrum of the acceleration and position noise are given by 

\begin{equation}
	\begin{split}
		S_p (f) = & N_p \left[1 + \left(\frac{2 \, \text{mHz}}{f}\right)^4 \right] \text{Hz}^{-1}, \\
		S_a (f) = &  \left[1 + \left(\frac{0.4 \, \text{mHz}}{f} \right)^2\right] \left[1 + \left(\frac{f}{8 \, \text{mHz}}  \right)^4 \right] \\ & \times \frac{N_a}{(2 \pi f)^4 } \text{Hz}^{-1}
	\end{split}
	\label{Eq:Instr_noise}
\end{equation}
The noise levels can be set by the end-user by modifying $N_p$ and $N_a$, but the current implementation assumes that they are the same in all satellite links. In this paper we set {$N_p = 9 \times 10^{-42} \text{ and } N_a =  3.6 \times 10^{-49} ~ \text{Hz}^{-4}$} to match the instrumental noise levels described in the LISA proposal. The code also implements time-delay interferometry (TDI) with Michelson, $X-Y-Z$ and $A-E-T$~\citep{Tinto:2004wu, Adams:2010vc} channels and heliocentric rigid-body orbits of the LISA satellites. This is implemented in an adiabatic manner by modeling the satellites to be stationary for small segments of time ($\ll 1 \text{ year}$) both for signal simulation and recovery, and allowing them to move between the time segments. The current implementation of orbits neglects the differential time delay for laser light on the round trip between two satellites -- i.e the travel time difference from satellite A to B and from B to A -- and also neglects their breathing modes. The satellites' orbital motion is especially important in partially breaking the degeneracies of the antenna patterns when detecting an anisotropic SGCN.

Finally, the BLIP pipeline is built to facilitate Bayesian inference and supports emcee~\citep{Foreman_Mackey_2013} and dynesty~\citep{Speagle_2019} samplers. All results in this paper were made through the dynesty sampler. The \cg coefficients are implemented with the help of the Wigner module of SymPy~\citep{Meurer:2017}

\begin{figure*}
\centering
\includegraphics[width=7.4in]{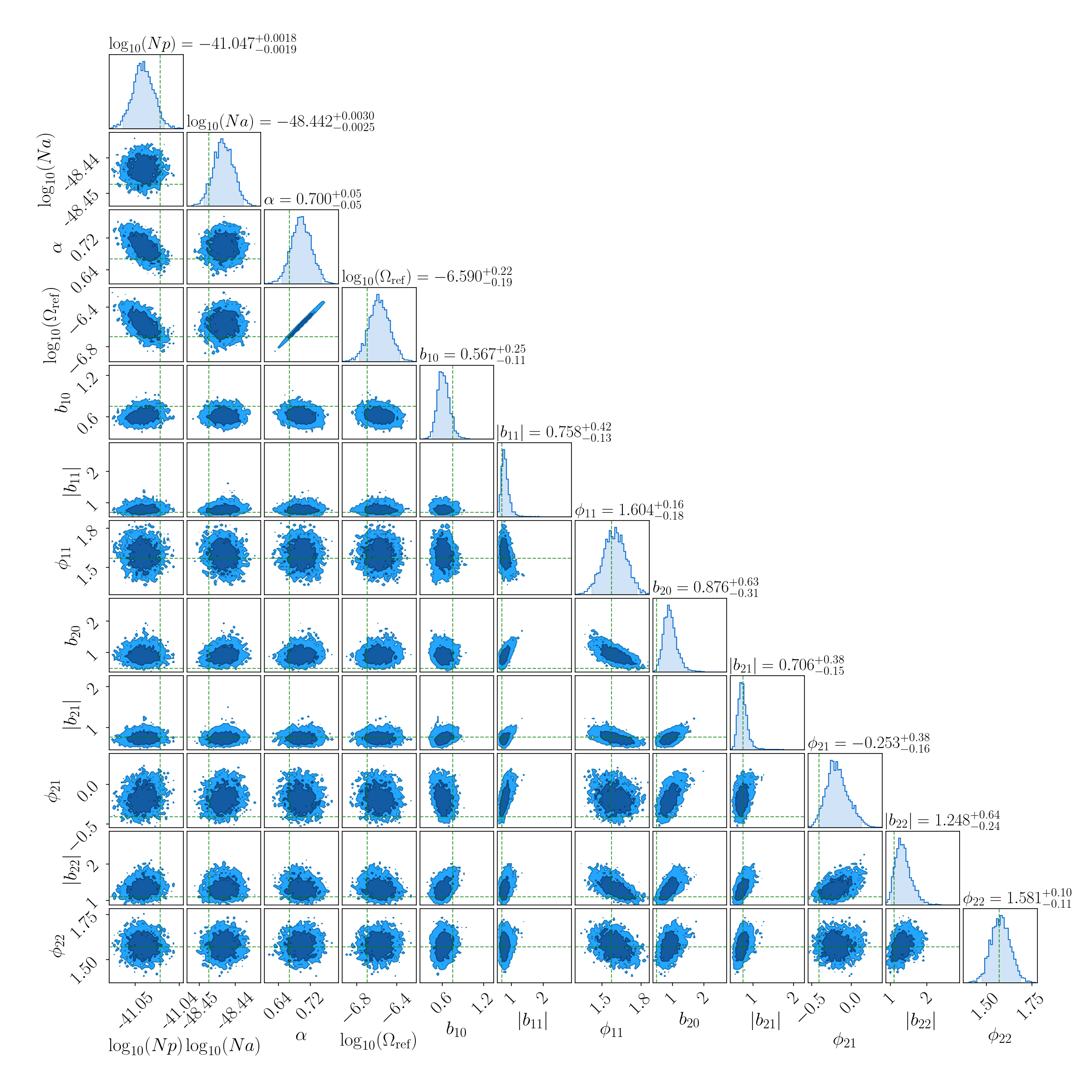}
 \caption{Posteriors corresponding to skymaps in Fig.~\ref{Fig:skymaps} with a signal amplitude $\Omega_{\text{ref}} = 2 \times 10^{-7}$ for the duration of $1$ year. This corresponds to a single channel theoretical SNR $\approx 149$. The shaded region in the one-dimensional posteriors are $95 \%$ confidence levels while the light and dark regions in the two-dimensional posteriors are one and two sigma confidence levels respectively. The parameters $N_p$ and $N_a$ are the posterior measurements for the position and acceleration noises respectively using the functional forms described in Eq.~\ref{Eq:Instr_noise}. The parameters $\alpha$ and $\Omega_{\text{ref}}$ measure the spectral shape of the SGCN while the rest of the parameters are measurements of $\blm$'s which describe the distribution of GW power on the sky. The dashed green lines are the true values of these parameters used when simulating the data. }
 \label{Fig:posteriors_1yr}
\end{figure*}

\subsection{Analysis configuration}
\label{Sec:config}

For the remainder of this paper we will use $X-Y-Z$ TDI channels with a rigid body orbiting configuration of LISA. To analyze the simulated data, we employ Fourier transforms with a duration of $T_{\text{seg}} = 1 \times 10^5 $ s with the aforementioned adiabatic approximation within each segment. The sampling frequency of the data is $f_s = 0.25 $ Hz. \redtext{We only consider the Fourier components to be between $f_{\text{min}} = 2 \times 10^{-4}$ Hz and $f_{\text{max}} = 2 \times 10^{-2}$ Hz. It is also desirable for computational purposes to  approximate the covariance matrix of the data in the frequency-time analysis to be diagonal across frequency and time. For this to be true, we require the auto-correlation time scale to be much smaller than $T_{\text{seg}}$~\citep{Banagiri:2019lon}. The value of  $T_{\text{seg}} = 1 \times 10^5 $ s  is thus chosen as a compromise between the auto-correlation time scale of LISA, which is $\sim 10^4$ s for this band, and the motion of the satellites.} Each time segment is Hann-windowed before being Fourier transformed. 

\subsection{Likelihood function}
\label{Sec:like}
 The power spectral density (PSD) and the cross-spectral density (CSD) of the data are combinations of the GW power from the SGCN and of the instrumental noise power~\citep{Cornish:2013nma}. Assuming that both are Gaussian, the Fourier domain likelihood is based on the multi-dimensional complex Gaussian distribution~\citep{Adams:2010vc}:

\begin{equation}
	\begin{split}
	\mathcal{L}(\tilde{d} | N_p, N_a, \Omega_{\rm ref}, \alpha, \{\blm\}) & =  \prod_{t, f} \frac{1}{2 \pi T_{\text{seg}} |C(t, f) |} \\  & \times  \exp \left( - \frac{2 \, \tilde{d}^*_{t, f} \, C(t, f)^{-1} \, \tilde{d}_{t, f}}{T_{\text{seg}}} \right). 
	\label{Eq:power_likelihood} 
	\end{split}
\end{equation} 
Here $\tilde{d}_{t, f} = [ \tilde{d}_X (t, f), \tilde{d}_Y(t, f), \tilde{d}_Z (t, f)   ]$ is the array of data in the Fourier domain for the three channels measured in the time segment labeled by $t$ and at frequency $f$. As previously mentioned, the data is Fourier transformed in segments of duration $T_{\text{seg}}$, and the product is across all frequency bins and time-segments. Here $C(t, f)$ is the $3 \times 3$ covariance matrix across the three channels, and as seen in Eqs.~\ref{Eq:Autocorr} and \ref{Eq:Crosscorr}, its elements are the sum of the signal spectral densities (defined in Eq.~\ref{Eq:PSD_sgwb}) and the instrumental noise spectral densities $S^n_{IJ} (f)$. We follow the derivation in~\cite{Adams:2010vc} of the actual expressions for the noise spectral densities. \redtext{Appendix~\ref{Sec:appx} goes into more detail and an explicit form of the covariance matrix in terms of the noise and signal spectral densities is given.} 

The term $|C (t, f) |$ is the determinant of the covariance matrix. In many applications of Bayesian inference to GW data, one usually can ignore the overall normalization because it is constant and model independent. \redtext{This is not true in this case, because the model parameters enter the likelihood through the covariance matrix.} Correctly modeling the normalization is thus essential to the the problem of Bayesian inference of the SGCN. \redtext{The matrix inverse is calculated numerically.}    

\begin{figure*}
\centering
\includegraphics[width=7.5in]{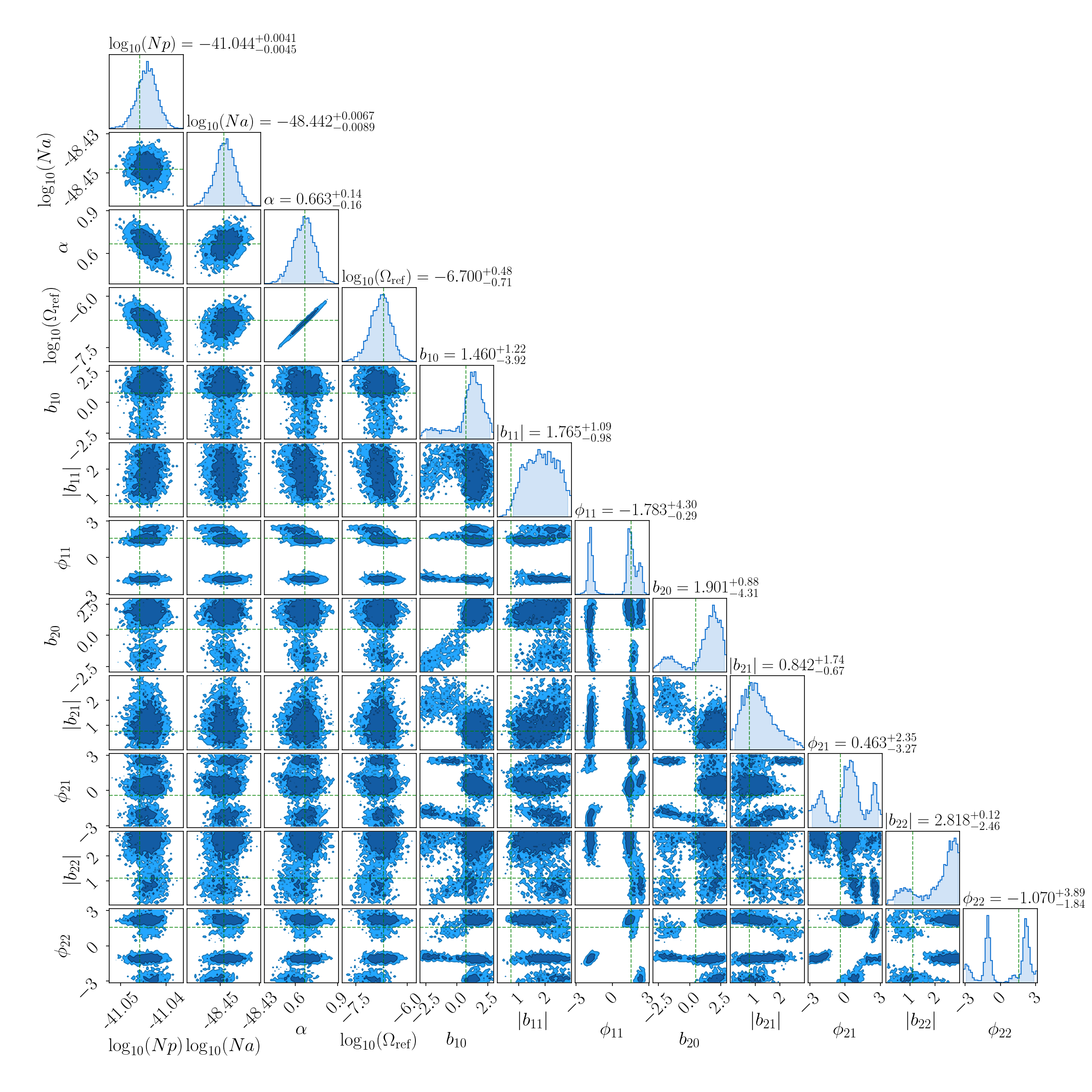}
 \caption{Posteriors for an analysis with a signal amplitude $\Omega_{\text{ref}} = 2 \times 10^{-7}$ for the duration of $2$ months. This corresponds to a single channel theoretical SNR $\approx 59$. The shaded region in the one-dimensional posteriors are $95 \%$ confidence levels while the light and dark regions in the two-dimensional posteriors are one and two sigma confidence levels respectively. The bimodalities seen in the one-dimensional posteriors are due to the parity symmetries described in Sec.~\ref{Sec:CG_decomp} which are only approximately broken in the limit of a weak signal or short integration time. With a stronger signal, the breaking of the symmetry becomes more complete and the degenerate modes go away as the sampler finds the right mode, as can be seen in the posterior for the $1$ year run (Fig. \ref{Fig:posteriors_1yr}).   }
 \label{Fig:posteriors_2mn}
\end{figure*}

\begin{figure*}
\subfigure[]
{
    \scalebox{0.4}{\includegraphics{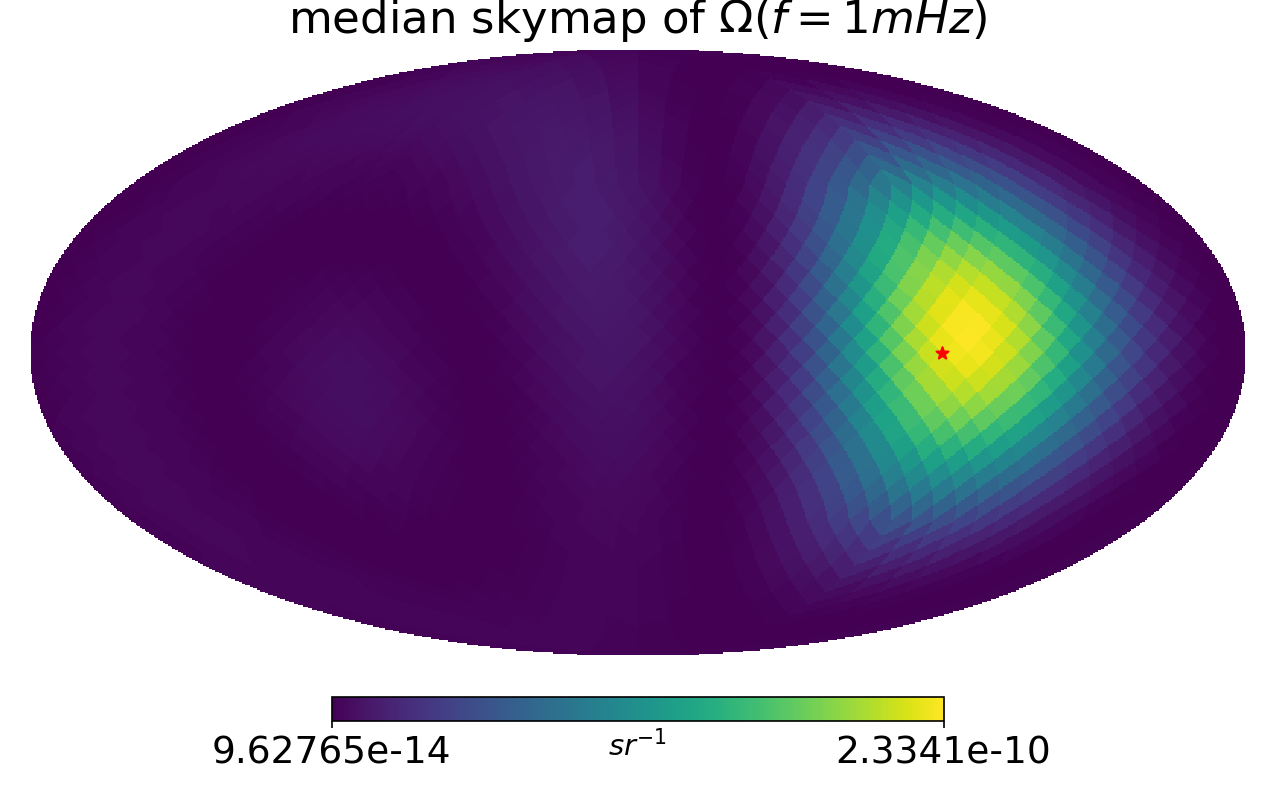}}
}
\subfigure[]
{
    \scalebox{0.4}{\includegraphics{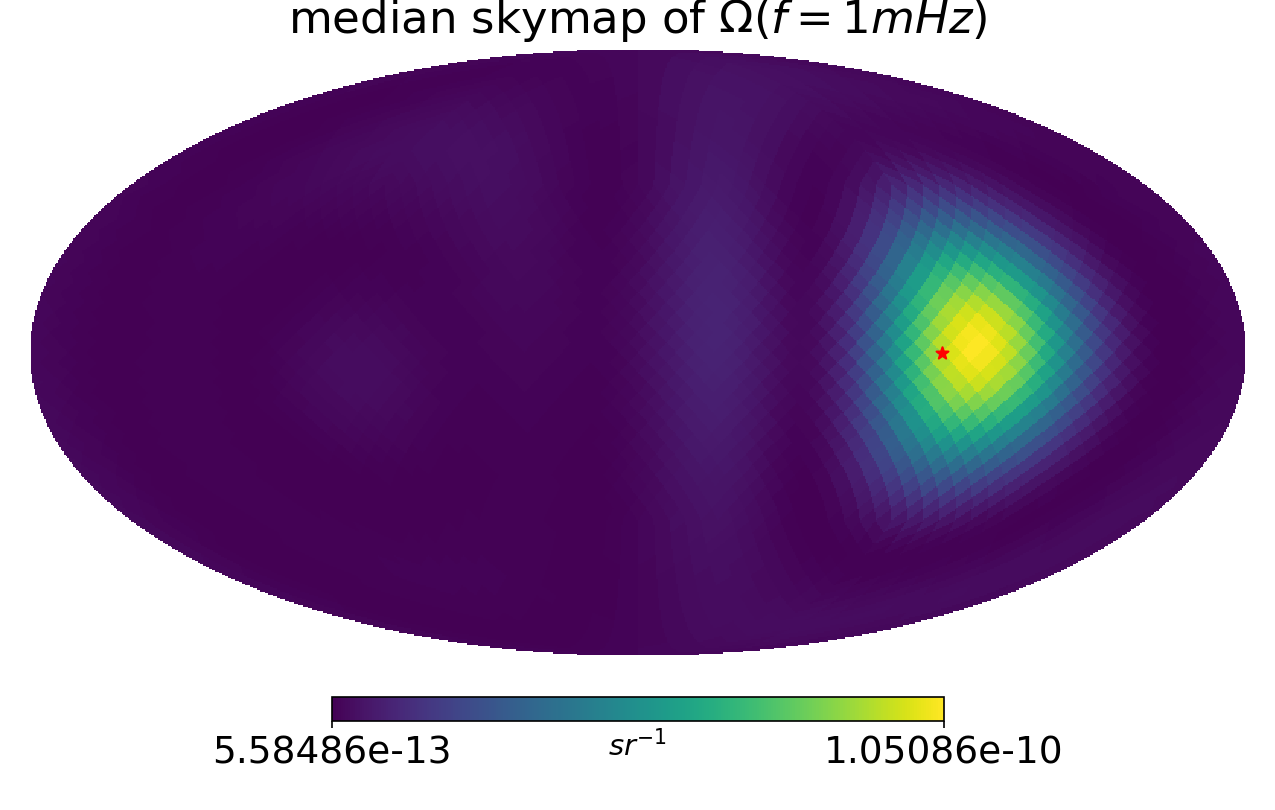}}
}
 \caption{Recovered skymaps for simulated localized sources using three months of simulated data. The red star indicates the true position of the source. The map on the left uses a cutoff of  $\ell_{\text{max}}^a=4$ while the map on the right uses $\ell_{\text{max}}^a=6$. Consequently, the power is smeared to a larger extent on the sky for the former compared to the latter. }
 \label{Fig:point}
\end{figure*}

\section{Simulation and detection}
\label{Sec:Sim_Det}

\subsection{Validation}
\label{Sec:validation}

We first validate the \cg technique by recovering an ad hoc distribution of power, simulated with a power-law spectral-index SGCN with {$\ell_{\text{max}}^a = 4$}. The  simulated $\{\blm\}$ coefficients are~\footnote{in HEALPix order} {($1.0, \, 0.75, \, 0.5, \, 0.7j, \, 0.7-0.3j, \, 1.1j )$} which yield the GW power distribution on the sky shown in Fig.~\ref{Fig:skymaps}. These values are chosen only to validate the ability of the algorithm to recover an arbitrary distribution of power on the sky. The spectral index of the power law is chosen to be consistent with binary inspiral at $\alpha = 2/3$ with {$\Omega (f = 25 \text{Hz}) = 2 \times 10^{-7}$}. 

The spherical harmonic coefficients $\{\blm\}$ are complex if $m\neq0$ and thus have two degrees of freedom. We parameterize them by their amplitude $|\blm|$ and phase $\phi_{\ell, m}$. We set uniform priors between $[0, 3]$ on the amplitude and uniform priors between $[-\pi, \pi]$ on the phase. For the modes with $m = 0$, i.e $b_{\ell, 0}$'s which are real, we set uniform priors between $[-3, 3]$. \redtext{To examine the support of this prior, we draw many samples from this prior distribution. Each sample corresponds to a unique distribution on the sky. Fig.~\ref{Fig:prior_skymap} shows the 2.5 and 97.5 percentile values of the sky-distribution of these samples at each pixel, without the factor of $\Omega (f)$. A HEALPix map with $n_{\text{side} = 6}$ is used. The range of these values shows the range of support per pixel from the prior. Note that while this helps with visualizing the prior range, this is not a perfect measure because the distributions across pixels are not independent of each other.  }

The variance of the prior sky map due to these choices of priors is shown in Fig.~\ref{Fig:prior_skymap}. This shows broad support for many modes and implies that the priors are not peaked at any particular region in the space.

We use the single channel signal-to-noise ratio (SNR) as a metric to characterize the strength of the signal, defined as

\begin{equation}
	\text{SNR} = \sqrt{T_{\text{seg}} \sum_{t, f} \delta f \, \frac{S^{GW}_{XX} (t, f) }{S^n_{XX} (f)} }
\end{equation}
where the summation is over time and frequency band used for analyzing the data. Note that while the SNR is a metric of the strength of the SGCN, \redtext{for a non-isotropic distribution of the SGCN, it does not grow as $\sim T^{1/2} $ as would be expected in the isotropic case. This is because the sensitivity to different directions change with time as LISA rotates in its orbit around the Sun. Hence, the GW power from an anisotropic SGCN will also exhibit a time modulation with a periodicity of a year.}

We ran two separate analyses with a duration of one year and two months. Figures~\ref{Fig:posteriors_1yr} and~\ref{Fig:posteriors_2mn} show the posterior corner plots for the former and the latter respectively. We witness bimodalities in the posteriors for the $\blm$ in the two-month run, which are related to the parity symmetry described in Sec.~\ref{Sec:CG_decomp}. While our choice of fixing $b_{0, 0}=1$ breaks this symmetry in principle, in practice, this breaking is only approximate and can fail in the limit of a weak signal or a small amount of observation time. This is because while two modes may be related by a sign change due to parity, we are limited by the ability of LISA to resolve this relative sign (with respect to $b_{0, 0}$). However, with enough time or a stronger signal, the breaking of the symmetry becomes complete, and the sampler finds the right mode as seen in the posterior for the one year run. The  injected skymap, and \redtext{the recovered median posterior skymaps for the two month and one year runs} are shown in Fig.~\ref{Fig:skymaps}. The median posterior map is the skymap generated from median values of posterior samples.

\redtext{To quantify the recovery of the angular power distribution, we use coherence as a measure of the similarity of the injected and recovered posterior skymaps~\citep{Romano:2015uma, Yang:2020usq, Taylor:2020zpk}. First, we transform the skymaps into a pixel basis on the sky using healpy. Since GW power is non-negative, we define excess power in pixel $i$ with respect to the mean as,} 

\begin{equation}
	\Delta \Omega^i = \Omega^i - \langle \Omega \rangle,
\end{equation}
\redtext{where $\langle \Omega \rangle$ is the average power across all pixels. The coherence $\Gamma$ between the injected skymap $ \Omega^i_{\text{inj} }$ and the recovered skymap $ \Omega^i_{\text{rec} }$ is then defined as}

\begin{equation}
	\Gamma = \frac{\sum_i \Delta \Omega^i_{\text{inj}} \, \Delta \Omega^i_{\text{rec}}     }{\sqrt{ \sum_i  (\Delta \Omega^i_{\text{inj}} )^2 } \sqrt{ \sum_i  (\Delta \Omega^i_{\text{rec}} )^2 } }, 
\end{equation}
\redtext{Defined in this way, the coherence is a real number between -1 to 1 as usual. Since we have the entire posterior of sky distributions, we can also measure the confidence of the coherence. The median and 95 \% quantiles on the coherence for the two months run is $ 0.426^{+0.428}_{-0.516}$. The median and 95 \% quantiles of the coherence for the one-year run is $ 0.956^{+0.032}_{-0.064}$. The improvement from the one-year run compared to the two-month run is evident not just by the better median coherence but also by much narrower quantiles of uncertainty. }

\subsection{Localized sources}
\label{Sec:point}

Next, we test the ability to recover signals from localized sources of GW power, often referred to as point sources in the literature. Examples of such sources could be combined gravitational radiation from a large globular cluster or a nearby galaxy. The GW power from these sources would be an incoherent superposition of all the GWs emitted by individual sources within them. Such sources are generally localized to much smaller angular scales than is possible to resolve with LISA. Thus when we map them with spherical harmonic methods, the power is usually smeared on larger angular scale as determined by the strength of the signal, the integration time and the $\ell_{\text{max}}^a$ scale where we cutoff the expansion. The dependence on the latter is demonstrated in Fig.~\ref{Fig:point} where three months of simulated data containing a point source signal is analyzed using $\ell_{\text{max}}^a = 4~\text{and}~6$, with the same $\Omega (f)$ in both cases. As expected the smearing of the power is smaller in the case of $\ell_{\text{max}}^a = 6$. In both cases, we still assumed $\alpha = 2/3$ power-law spectrum.

\section{Galactic Foreground}
\label{Sec:Foreground}

The galactic white dwarf foreground is expected to be one of the primary sources of confusion noise seen by LISA. Due to its strength and strong anisotropy, characterizing this foreground will be a primary application of the method described here. Accurately mapping the foreground will also be important for searches of resolvable sources as it is a major source of noise. The confusion noise in the detectors due to the galactic foreground will exhibit a time modulation tied to its anisotropy as LISA orbits the sun. 

We simulate the galactic foreground by modeling it as stochastic colored-Gaussian noise with a power-law spectrum given by Eq.~\ref{Eq:Power_law}. Following~\cite{Breivik:2019lmt, Breivik:2019oar} and~\cite{McMillan:2011}, we model the spatial distribution in the galaxy using a disk+bulge model. However, instead of simulating an ensemble of individual DWD systems, we instead distribute the DWD density in the galactocentric frame using this model. The disk is modeled as,

\begin{equation}
	\rho_{d}(r, z) \propto \exp \left ( - r/ r_h \right) \exp \left( -z /z_h \right),
\end{equation}
where the radial scale height is assumed to be $r_h = 2.9$ kpc, and the vertical scale height $z_h = 0.3$ kpc following the thin disk model of~\cite{Breivik:2019lmt}. The galactic bulge is modeled to be azimuthally symmetric,

\begin{equation}
	 \rho_b (r) \propto \frac{\exp \left ( - (r / r_{\text{cut}} )^2 \, \right)  }{ (1 + r'/r_0)^{\gamma} }, 
\end{equation}
where $\gamma = 1.8$, $r_0 = 0.075$ kpc, $r_{\text{cut}} = 2.1$ kpc, $r' = \sqrt{r^2 + (z/q)^2}$ and $q = 0.5$. 

\begin{figure}
\subfigure[]
{
    \scalebox{0.4}{\includegraphics{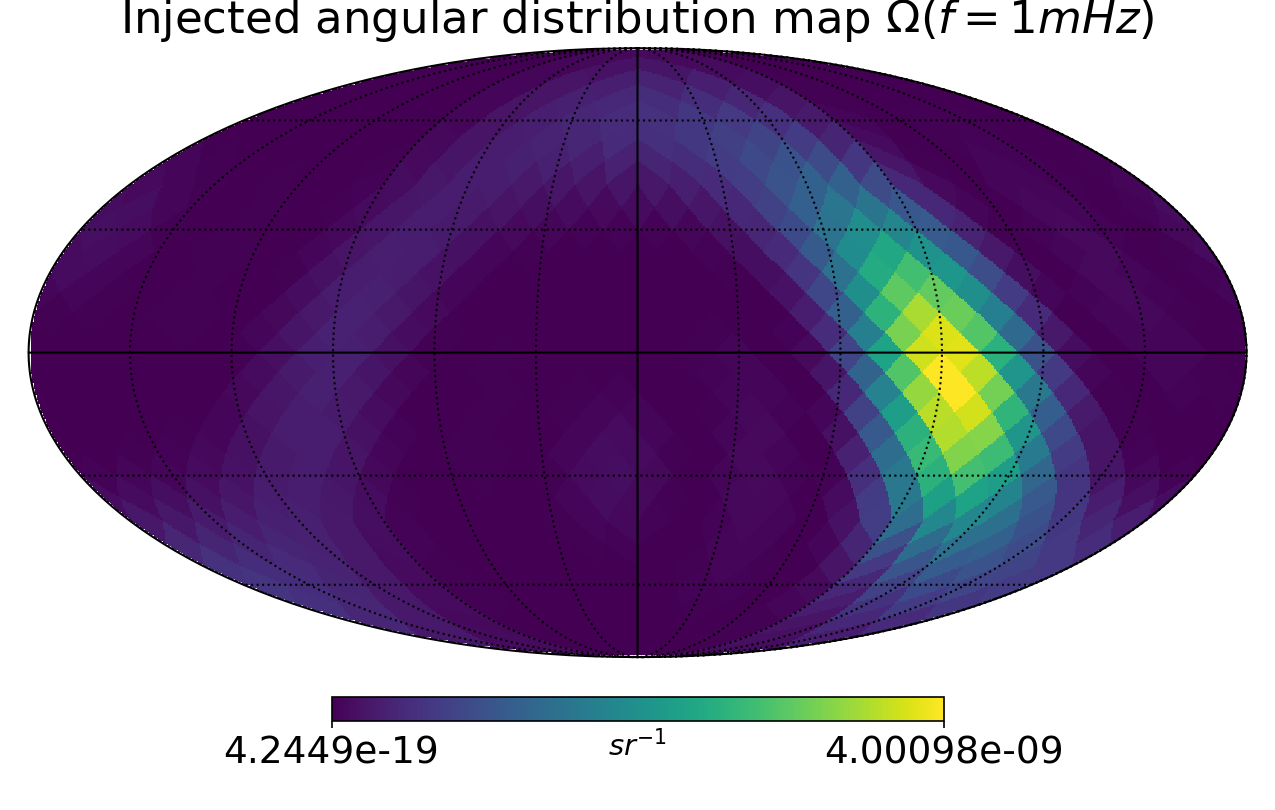}}
}
\subfigure[]
{
    \scalebox{0.4}{\includegraphics{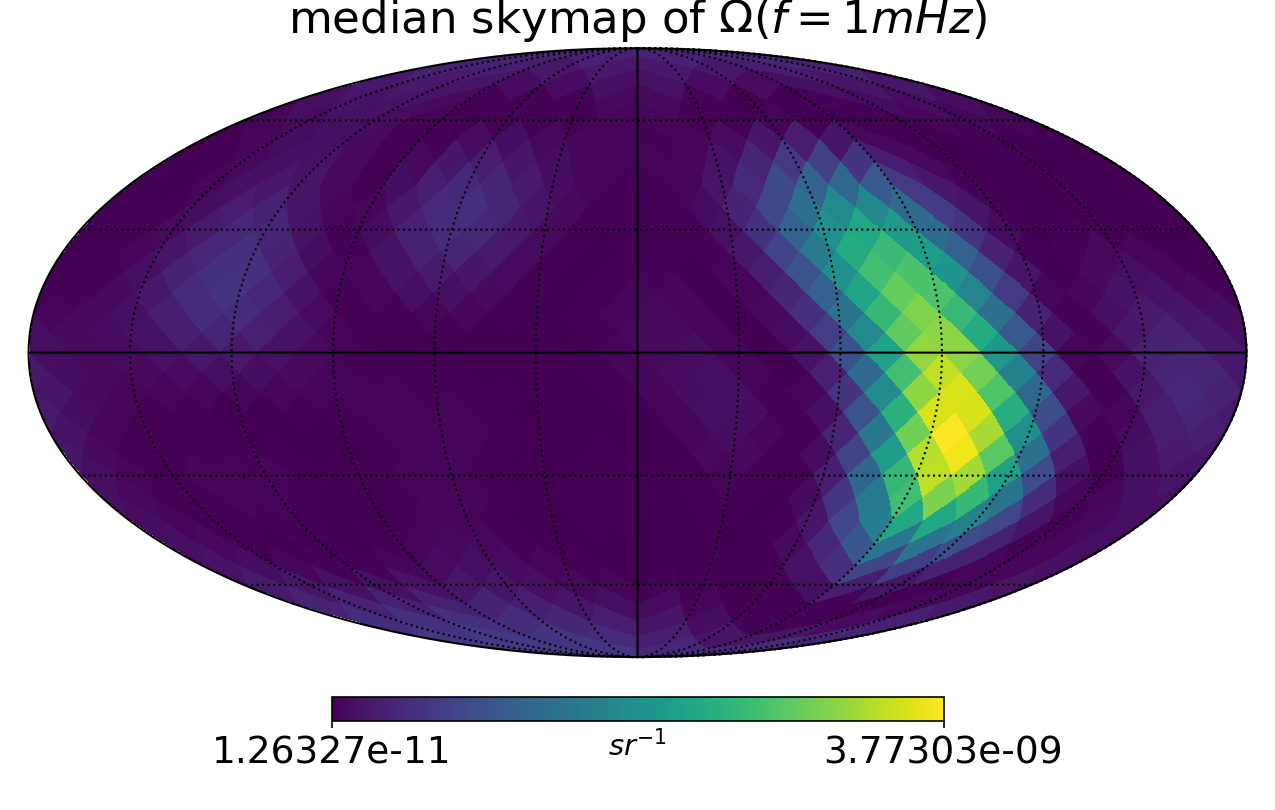}}
}
 \caption{Skymaps for the simulation and recovery of the galactic DWD foregrounds described in Sec.\ref{Sec:Foreground} with one year of data. The top panel shows the simulated skymap while the bottom shows posterior median recovery skymap. Both maps show the distribution of $\Omega(f = 1 \times 10^{-3}$ Hz) in the solar system barycentric frame. The bright spots in the map corresponds to the galactic central bulge.}
 \label{Fig:foreground_maps}
\end{figure}

The distribution is then transformed to the Solar System barycenter coordinates. The DWD density at each point in this grid is modulated by the distance to it from Earth and the grid is projected on to the sky to produce an unnormalized distribution on the sky. The skymap is then normalized to one and multiplied by the power spectrum of the foreground, modeled as an $\alpha = 2/3$ power law, producing the simulated GW power skymap seen on the top panel of Fig.~\ref{Fig:foreground_maps}. The GW power distribution is then convolved with the time-dependent LISA detector response to calculate the GW power in the detector from which the simulated time-domain data is generated.

Here, we have simulated a year of galactic foreground data in the manner described above, which we then analyzed using the \cg method with $\ell_{max}^a = 4$. In the analysis, we have used $f_{\text{max}}$ of $10^{-3}$ Hz to simulate the drop in the foreground due to mass transfer DWDs in a simplistic manner. The bottom panel of Fig.~\ref{Fig:foreground_maps} shows the recovered skymap. We see that the overall shape is qualitatively well recovered, especially the galactic bulge. \redtext{We also compute the coherence between the injected and recovered map. The median and 95 \% quantiles on the coherence for the foreground analysis is $ 0.924^{+0.018}_{-0.024}$.}  This demonstrates the \cg method's efficacy in mapping the galactic DWD foreground. We leave a more sophistical analysis of recovering the foreground's actual shape, and the exact frequency cut off caused by mass transfer to a future paper.

\section{Discussion \& Conclusion}
\label{Sec:Conclusion}

We have developed a Bayesian mapping algorithm using the spherical harmonic basis that can optimally recover any arbitrary distribution of GW power on the sky using LISA data, while imposing the physical constraint that the GW power is non-negative in all directions on the sky. This method was validated through a series of end-to-end simulations of different types of SGCNs.

While this paper developed mapping tools and validates them, there are several directions to push this forward and apply them for astrophysical use. The galactic foreground simulated in this paper is ultimately simplistic in that it assumes that it is Gaussian and has the spectral shape of a power law with $\alpha = 2/3$. It will be essential to relax these assumptions and validate this method on a realistic simulated foreground formed from catalogs of DWDs generated with population synthesis codes. This might also require relaxing the assumptions of Gaussianity. Additionally, as \citet{Breivik:2019oar} shows, mapping the foreground on the sky can help constrain galactic structure in a manner complementary to resolvable DWDs while also probing the stellar evolution history of the galactic white dwarfs. It would be interesting to study what mapping the foreground could tell us about the astrophysics of galactic foreground with the optimal analysis developed in this work and using realistic simulations. One can also expand this to study the astrophysics of the galactic DWD population in a more model-independent manner by constraining the properties of the foreground in narrow frequency bands.

It will also be essential to study how the algorithm's angular sensitivity, characterized by the $\ell_{\text{max}}^a$ parameter, will scale with the strength of the stochastic confusion noise and the duration of the analysis. \redtext{A data-centric way would be to determine the optimal $\ell_{\text{max}}^a$ directly from the data based on statistical considerations, for example, using Bayesian model selection. We can also make $\ell_{\text{max}}^a$ an independent parameter that is directly fit by the data. Since the dimensionality of the spherical harmonic parameter space itself depends on $\ell_{\text{max}}^a$, such an analysis would need to use Reversible-Jump Markov-chain Monte Carlo approaches similar to those developed in \citet{Cornish:2014kda}}.

Finally, an important application would be a joint analysis with both resolvable signals and the galactic DWD foreground. One of the core strengths of mapping the foreground in a Bayesian manner is the possibility of simultaneous inference of resolvable signals along with the galactic foreground. This will allow an unbiased estimation of the properties of the resolvable signals while accounting for the temporal modulation of the noise due to the galactic foreground. \redtext{One can also imagine multiple SGCNs, arising from different sources that might be separable if they have different spectral shape and angular structure (for e.g.~\cite{Adams:2013qma}). }

\section*{Acknowledgments}
We are grateful to Katelyn Breivik for helpful discussions about the galactic white dwarf foreground, and to Ethan Payne for discussion about the \cg decomposition. We are thankful to Arianna Renzini for useful discussion about using the spherical harmonic basis with LISA. \redtext{We also thank the anonymous referee for several insightful suggestions that made the paper better.} 

All corner plots were made with ChainConsumer~\citep{ChainConsumer}. \redtext{The sky maps are made with the healpy python package using the HEALPix library~\citep{healpy2, Zonca2019}}. The authors acknowledge the Minnesota Supercomputing Institute (MSI) at the University of Minnesota for providing computational resources that contributed to the research reported within this paper. This work is supported by the NASA grant 80NSSC19K0318. SB acknowledges support by the Doctoral Dissertation Fellowship at the University of Minnesota. AC acknowledges support by the National Science Foundation under grant No. 1922512. SRT acknowledge support from NSF AST-2007993, and a Dean’s Faculty Fellowship of Vanderbilt University’s College of Arts \& Science.

\section*{Data Availability statement}

No real data was generated or analyzed in this paper. The simulated mock data was both generated and analyzed with the BLIP code available at \url{https://github.com/sharanbngr/blip}.

\appendix

\section{The covariance matrix in the presence of an SGCN}
\label{Sec:appx}

\redtext{As described by Eqs.~\ref{Eq:Autocorr} and \ref{Eq:Crosscorr}, the cross-spectral density between channel $I$ and channel $J$ can be written as a linear combination of the instrumental noise power and the GW power due to the SGCN,}

\begin{equation}
	S_{IJ}(t, f) = S^n_{IJ} (f) + S^{GW}_{IJ} (t, f). 
\end{equation} 
\redtext{As shown by Eq.~\ref{Eq:PSD_sgwb}, the GW power can be written as a summation over the response function to spherical harmonic modes,} 

\begin{equation}
	S_{IJ}^{\text{GW}} (f, t) = \frac{3 H^2_0}{2 \pi^2 f^3} \frac{\Omega (f) }{\sqrt{4 \pi a_{0, 0}}} \sum_{\ell, m} \alm \mathcal{R}_{\ell, m}^{IJ} (f, t) .
\end{equation}

\redtext{We assume a strictly equal-arm LISA configuration with the same noise levels in all links, for which the noise auto-power of the X-Y-Z TDI channels can be written in terms of acceleration and position noise, defined in Eq~\ref{Eq:Instr_noise}, as (see the appendix of ~\cite{Adams:2010vc}):} 

\begin{equation}
	S^n_{II} = 16 \sin^2 \left(\frac{f}{f_*} \right) \left[S_p (f) + 2 S_a (f) \left( 1 + \cos^2 \left( \frac{f}{f_*} \right) \right) \right]. 
\end{equation}
\redtext{Similarly, the cross-power between two channels $I, J$ where $I \neq J$ is,} 

\begin{equation}
		S^n_{IJ} = 4 \sin^2 \left(\frac{f}{f_*} \right) \cos \left( \frac{f}{f_*} \right) \left[ -2S_p (f) - 8 S_a (f) \right] . 
\end{equation}
\redtext{Note that because the Michelson channels share arms, the noise between channels is not independent of each other, as evident by the non-zero cross-power. This requires us to use a full covariance matrix for statistical analysis through the likelihood of Eq.~\ref{Eq:power_likelihood}; treating the three channels independently would lead to biased inferences.} 

\redtext{ We can now write down the elements of the covariance matrix in the presence of an SGCN. The diagonal elements in the X-Y-Z combination are given by}

\begin{equation}
	\begin{split}
	C_{II} (t, f) = & \frac{3 H^2_0}{2 \pi^2 f^3}  \frac{\Omega (f) }{\sqrt{4 \pi a_{0, 0}}} \sum_{\ell, m} \alm \mathcal{R}_{\ell, m}^{II} (f, t) \\  +  & \, 16 \sin^2 \left(\frac{f}{f_*} \right)  \left[S_p (f) + 2 S_a (f) \left( 1 + \cos^2 \left( \frac{f}{f_*} \right) \right) \right], 
	\end{split}
\end{equation}
\redtext{whereas the off-diagonal elements ($I \neq J$) are given by , }

\begin{equation}
		\begin{split}
	C_{IJ} (t, f) = & \frac{3 H^2_0}{2 \pi^2 f^3}  \frac{\Omega (f) }{\sqrt{4 \pi a_{0, 0}}} \sum_{\ell, m} \alm \mathcal{R}_{\ell, m}^{IJ} (f, t) \\  +  & \, 4 \sin^2 \left(\frac{f}{f_*} \right) \cos \left( \frac{f}{f_*} \right) \left[ -2S_p (f) - 8 S_a (f) \right] .  
	\end{split}
\end{equation}
\redtext{We now see how the independent parameters that are sampled over enter the covariance matrix. The noise parameters $N_p$ and $N_a$ enter through the acceleration and position power spectra as seen in Eq.~\ref{Eq:Instr_noise}. The $\Omega_{\text{ref}}, \alpha$ and spherical harmonic $\blm$ parameters that describe the SGCN enter through the GW power; the latter being related to the $\alm$ parameters through Eq.~\ref{Eq:alm_2_blm}. }

\bibliography{sph}

\begin{thebibliography}{}
\makeatletter
\relax
\def\mn@urlcharsother{\let\do\@makeother \do\$\do\&\do\#\do\^\do\_\do\%\do\~}
\def\mn@doi{\begingroup\mn@urlcharsother \@ifnextchar [ {\mn@doi@}
  {\mn@doi@[]}}
\def\mn@doi@[#1]#2{\def\@tempa{#1}\ifx\@tempa\@empty \href
  {http://dx.doi.org/#2} {doi:#2}\else \href {http://dx.doi.org/#2} {#1}\fi
  \endgroup}
\def\mn@eprint#1#2{\mn@eprint@#1:#2::\@nil}
\def\mn@eprint@arXiv#1{\href {http://arxiv.org/abs/#1} {{\tt arXiv:#1}}}
\def\mn@eprint@dblp#1{\href {http://dblp.uni-trier.de/rec/bibtex/#1.xml}
  {dblp:#1}}
\def\mn@eprint@#1:#2:#3:#4\@nil{\def\@tempa {#1}\def\@tempb {#2}\def\@tempc
  {#3}\ifx \@tempc \@empty \let \@tempc \@tempb \let \@tempb \@tempa \fi \ifx
  \@tempb \@empty \def\@tempb {arXiv}\fi \@ifundefined
  {mn@eprint@\@tempb}{\@tempb:\@tempc}{\expandafter \expandafter \csname
  mn@eprint@\@tempb\endcsname \expandafter{\@tempc}}}

\bibitem[\protect\citeauthoryear{Aasi et~al.}{Aasi
  et~al.}{2015}]{TheLIGOScientific:2014jea}
Aasi J.,  et~al., 2015, \mn@doi [Class. Quant. Grav.]
  {10.1088/0264-9381/32/7/074001}, 32, 074001

\bibitem[\protect\citeauthoryear{Abbott et~al.}{Abbott
  et~al.}{2021a}]{Abbott:2021xxi}
Abbott R.,  et~al., 2021a, \mn@doi [Phys. Rev. D]
  {10.1103/PhysRevD.104.022004}, 104, 022004

\bibitem[\protect\citeauthoryear{Abbott et~al.}{Abbott
  et~al.}{2021b}]{Abbott:2021jel}
Abbott R.,  et~al., 2021b, \mn@doi [Phys. Rev. D]
  {10.1103/PhysRevD.104.022005}, 104, 022005

\bibitem[\protect\citeauthoryear{Acernese et~al.}{Acernese
  et~al.}{2015}]{TheVirgo:2014hva}
Acernese F.,  et~al., 2015, \mn@doi [Class. Quant. Grav.]
  {10.1088/0264-9381/32/2/024001}, 32, 024001

\bibitem[\protect\citeauthoryear{Adams \& Cornish}{Adams \&
  Cornish}{2010}]{Adams:2010vc}
Adams M.~R.,  Cornish N.~J.,  2010, \mn@doi [Phys. Rev. D]
  {10.1103/PhysRevD.82.022002}, 82, 022002

\bibitem[\protect\citeauthoryear{Adams \& Cornish}{Adams \&
  Cornish}{2014}]{Adams:2013qma}
Adams M.~R.,  Cornish N.~J.,  2014, \mn@doi [Phys. Rev. D]
  {10.1103/PhysRevD.89.022001}, 89, 022001

\bibitem[\protect\citeauthoryear{Allen \& Romano}{Allen \&
  Romano}{1999}]{Allen:1997ad}
Allen B.,  Romano J.~D.,  1999, \mn@doi [Phys. Rev. D]
  {10.1103/PhysRevD.59.102001}, 59, 102001

\bibitem[\protect\citeauthoryear{Alonso, Contaldi, Cusin, Ferreira  \&
  Renzini}{Alonso et~al.}{2020}]{Alonso:2020rar}
Alonso D.,  Contaldi C.~R.,  Cusin G.,  Ferreira P.~G.,   Renzini A.~I.,  2020,
  \mn@doi [Phys. Rev. D] {10.1103/PhysRevD.101.124048}, 101, 124048

\bibitem[\protect\citeauthoryear{Amaro-Seoane, Gair, Freitag, Coleman~Miller,
  Mandel, Cutler  \& Babak}{Amaro-Seoane et~al.}{2007}]{AmaroSeoane:2007aw}
Amaro-Seoane P.,  Gair J.~R.,  Freitag M.,  Coleman~Miller M.,  Mandel I.,
  Cutler C.~J.,   Babak S.,  2007, \mn@doi [Class. Quant. Grav.]
  {10.1088/0264-9381/24/17/R01}, 24, R113

\bibitem[\protect\citeauthoryear{Amaro-Seoane. et~al.}{Amaro-Seoane.
  et~al.}{2017}]{amaroseoane2017laser}
Amaro-Seoane. P.,  et~al., 2017, arXiv e-prints

\bibitem[\protect\citeauthoryear{Arfken}{Arfken}{2012}]{ArfkenGeorge1972MMfP}
Arfken G.,  2012, Mathematical Methods for Physicists.
Elsevier, \mn@doi{https://doi.org/10.1016/C2009-0-30629-7}

\bibitem[\protect\citeauthoryear{Arzoumanian et~al.}{Arzoumanian
  et~al.}{2020}]{Arzoumanian:2020vkk}
Arzoumanian Z.,  et~al., 2020, \mn@doi [Astrophys. J. Lett.]
  {10.3847/2041-8213/abd401}, 905, L34

\bibitem[\protect\citeauthoryear{Auclair et~al.}{Auclair
  et~al.}{2020}]{Auclair:2019wcv}
Auclair P.,  et~al., 2020, \mn@doi [JCAP] {10.1088/1475-7516/2020/04/034}, 04,
  034

\bibitem[\protect\citeauthoryear{Babak et~al.,}{Babak
  et~al.}{2017}]{Babak:2017tow}
Babak S.,  et~al., 2017, \mn@doi [Phys. Rev. D] {10.1103/PhysRevD.95.103012},
  95, 103012

\bibitem[\protect\citeauthoryear{Banagiri, Coughlin, Clark, Lasky, Bizouard,
  Talbot, Thrane  \& Mandic}{Banagiri et~al.}{2020}]{Banagiri:2019lon}
Banagiri S.,  Coughlin M.~W.,  Clark J.,  Lasky P.~D.,  Bizouard M.,  Talbot
  C.,  Thrane E.,   Mandic V.,  2020, \mn@doi [Mon. Not. Roy. Astron. Soc.]
  {10.1093/mnras/staa181}, 492, 4945

\bibitem[\protect\citeauthoryear{Barausse, Bellovary, Berti, Holley-Bockelmann,
  Farris, Sathyaprakash  \& Sesana}{Barausse et~al.}{2015}]{Barausse:2014oca}
Barausse E.,  Bellovary J.,  Berti E.,  Holley-Bockelmann K.,  Farris B.,
  Sathyaprakash B.,   Sesana A.,  2015, \mn@doi [J. Phys. Conf. Ser.]
  {10.1088/1742-6596/610/1/012001}, 610, 012001

\bibitem[\protect\citeauthoryear{Bartolo et~al.}{Bartolo
  et~al.}{2016}]{Bartolo:2016ami}
Bartolo N.,  et~al., 2016, \mn@doi [JCAP] {10.1088/1475-7516/2016/12/026}, 12,
  026

\bibitem[\protect\citeauthoryear{Bartolo et~al.,}{Bartolo
  et~al.}{2020}]{Bartolo:2019zvb}
Bartolo N.,  et~al., 2020, \mn@doi [JCAP] {10.1088/1475-7516/2020/02/028}, 02,
  028

\bibitem[\protect\citeauthoryear{Benacquista \& Holley-Bockelmann}{Benacquista
  \& Holley-Bockelmann}{2006}]{Benacquista:2005tm}
Benacquista M.,  Holley-Bockelmann K.,  2006, \mn@doi [Astrophys. J.]
  {10.1086/504024}, 645, 589

\bibitem[\protect\citeauthoryear{Brazier et~al.}{Brazier
  et~al.}{2019}]{Brazier:2019mmu}
Brazier A.,  et~al., 2019

\bibitem[\protect\citeauthoryear{Breivik et~al.}{Breivik
  et~al.}{2020a}]{Breivik:2019lmt}
Breivik K.,  et~al., 2020a, \mn@doi [Astrophys. J.] {10.3847/1538-4357/ab9d85},
  898, 71

\bibitem[\protect\citeauthoryear{Breivik, Mingarelli  \& Larson}{Breivik
  et~al.}{2020b}]{Breivik:2019oar}
Breivik K.,  Mingarelli C. M.~F.,   Larson S.~L.,  2020b, \mn@doi [Astrophys.
  J.] {10.3847/1538-4357/abab99}, 901, 4

\bibitem[\protect\citeauthoryear{Caprini et~al.}{Caprini
  et~al.}{2016}]{Caprini:2015zlo}
Caprini C.,  et~al., 2016, \mn@doi [JCAP] {10.1088/1475-7516/2016/04/001}, 04,
  001

\bibitem[\protect\citeauthoryear{Contaldi, Pieroni, Renzini, Cusin, Karnesis,
  Peloso, Ricciardone  \& Tasinato}{Contaldi et~al.}{2020}]{Contaldi:2020rht}
Contaldi C.~R.,  Pieroni M.,  Renzini A.~I.,  Cusin G.,  Karnesis N.,  Peloso
  M.,  Ricciardone A.,   Tasinato G.,  2020, \mn@doi [Phys. Rev. D]
  {10.1103/PhysRevD.102.043502}, 102, 043502

\bibitem[\protect\citeauthoryear{Cornish}{Cornish}{2001}]{Cornish:2001hg}
Cornish N.~J.,  2001, \mn@doi [Class. Quant. Grav.]
  {10.1088/0264-9381/18/20/307}, 18, 4277

\bibitem[\protect\citeauthoryear{Cornish}{Cornish}{2002a}]{Cornish:2002bh}
Cornish N.~J.,  2002a, \mn@doi [Class. Quant. Grav.]
  {10.1088/0264-9381/19/7/306}, 19, 1279

\bibitem[\protect\citeauthoryear{Cornish}{Cornish}{2002b}]{Cornish:2001bb}
Cornish N.~J.,  2002b, \mn@doi [Phys. Rev. D] {10.1103/PhysRevD.65.022004}, 65,
  022004

\bibitem[\protect\citeauthoryear{Cornish \& Larson}{Cornish \&
  Larson}{2001}]{Cornish:2001qi}
Cornish N.~J.,  Larson S.~L.,  2001, \mn@doi [Class. Quant. Grav.]
  {10.1088/0264-9381/18/17/308}, 18, 3473

\bibitem[\protect\citeauthoryear{Cornish \& Littenberg}{Cornish \&
  Littenberg}{2015}]{Cornish:2014kda}
Cornish N.~J.,  Littenberg T.~B.,  2015, \mn@doi [Class. Quant. Grav.]
  {10.1088/0264-9381/32/13/135012}, 32, 135012

\bibitem[\protect\citeauthoryear{Cornish \& Romano}{Cornish \&
  Romano}{2013}]{Cornish:2013nma}
Cornish N.~J.,  Romano J.~D.,  2013, \mn@doi [Phys. Rev. D]
  {10.1103/PhysRevD.87.122003}, 87, 122003

\bibitem[\protect\citeauthoryear{Cusin, Dvorkin, Pitrou  \& Uzan}{Cusin
  et~al.}{2018}]{Cusin:2018rsq}
Cusin G.,  Dvorkin I.,  Pitrou C.,   Uzan J.-P.,  2018, \mn@doi [Phys. Rev.
  Lett.] {10.1103/PhysRevLett.120.231101}, 120, 231101

\bibitem[\protect\citeauthoryear{Cusin, Dvorkin, Pitrou  \& Uzan}{Cusin
  et~al.}{2020}]{Cusin:2019jhg}
Cusin G.,  Dvorkin I.,  Pitrou C.,   Uzan J.-P.,  2020, \mn@doi [Mon. Not. Roy.
  Astron. Soc.] {10.1093/mnrasl/slz182}, 493, L1

\bibitem[\protect\citeauthoryear{Danielski, Korol, Tamanini  \&
  Rossi}{Danielski et~al.}{2019}]{Danielski:2019rvt}
Danielski C.,  Korol V.,  Tamanini N.,   Rossi E.,  2019, \mn@doi [Astron.
  Astrophys.] {10.1051/0004-6361/201936729}, 632, A113

\bibitem[\protect\citeauthoryear{Fitz~Axen, Banagiri, Matas, Caprini  \&
  Mandic}{Fitz~Axen et~al.}{2018}]{Axen:2018zvb}
Fitz~Axen M.,  Banagiri S.,  Matas A.,  Caprini C.,   Mandic V.,  2018, \mn@doi
  [Phys. Rev. D] {10.1103/PhysRevD.98.103508}, 98, 103508

\bibitem[\protect\citeauthoryear{Foreman-Mackey, Hogg, Lang  \&
  Goodman}{Foreman-Mackey et~al.}{2013}]{Foreman_Mackey_2013}
Foreman-Mackey D.,  Hogg D.~W.,  Lang D.,   Goodman J.,  2013, \mn@doi
  [Publications of the Astronomical Society of the Pacific] {10.1086/670067},
  125, 306

\bibitem[\protect\citeauthoryear{Gair, Babak, Sesana, Amaro-Seoane, Barausse,
  Berry, Berti  \& Sopuerta}{Gair et~al.}{2017}]{Gair:2017ynp}
Gair J.~R.,  Babak S.,  Sesana A.,  Amaro-Seoane P.,  Barausse E.,  Berry
  C.~P.,  Berti E.,   Sopuerta C.,  2017, \mn@doi [J. Phys. Conf. Ser.]
  {10.1088/1742-6596/840/1/012021}, 840, 012021

\bibitem[\protect\citeauthoryear{{G{\'o}rski}, {Hivon}, {Banday}, {Wandelt},
  {Hansen}, {Reinecke}  \& {Bartelmann}}{{G{\'o}rski} et~al.}{2005}]{healpy2}
{G{\'o}rski} K.~M.,  {Hivon} E.,  {Banday} A.~J.,  {Wandelt} B.~D.,  {Hansen}
  F.~K.,  {Reinecke} M.,   {Bartelmann} M.,  2005, \mn@doi [\apj]
  {10.1086/427976}, \href {http://adsabs.harvard.edu/abs/2005ApJ...622..759G}
  {622, 759}

\bibitem[\protect\citeauthoryear{{Hinton}}{{Hinton}}{2016}]{ChainConsumer}
{Hinton} S.~R.,  2016, \mn@doi [The Journal of Open Source Software]
  {10.21105/joss.00045}, \href
  {http://adsabs.harvard.edu/abs/2016JOSS....1...45H} {1, 00045}

\bibitem[\protect\citeauthoryear{Jenkins, Sakellariadou, Regimbau  \&
  Slezak}{Jenkins et~al.}{2018}]{Jenkins:2018uac}
Jenkins A.~C.,  Sakellariadou M.,  Regimbau T.,   Slezak E.,  2018, \mn@doi
  [Phys. Rev. D] {10.1103/PhysRevD.98.063501}, 98, 063501

\bibitem[\protect\citeauthoryear{Jenkins, Romano  \& Sakellariadou}{Jenkins
  et~al.}{2019}]{Jenkins:2019nks}
Jenkins A.~C.,  Romano J.~D.,   Sakellariadou M.,  2019, \mn@doi [Phys. Rev. D]
  {10.1103/PhysRevD.100.083501}, 100, 083501

\bibitem[\protect\citeauthoryear{Klein et~al.}{Klein
  et~al.}{2016}]{Klein:2015hvg}
Klein A.,  et~al., 2016, \mn@doi [Phys. Rev. D] {10.1103/PhysRevD.93.024003},
  93, 024003

\bibitem[\protect\citeauthoryear{Korol, Rossi, Groot, Nelemans, Toonen  \&
  Brown}{Korol et~al.}{2017}]{Korol:2017qcx}
Korol V.,  Rossi E.~M.,  Groot P.~J.,  Nelemans G.,  Toonen S.,   Brown A.~G.,
  2017, \mn@doi [Mon. Not. Roy. Astron. Soc.] {10.1093/mnras/stx1285}, 470,
  1894

\bibitem[\protect\citeauthoryear{Korol, Koop  \& Rossi}{Korol
  et~al.}{2018}]{Korol:2018ulo}
Korol V.,  Koop O.,   Rossi E.~M.,  2018, \mn@doi [Astrophys. J. Lett.]
  {10.3847/2041-8213/aae587}, 866, L20

\bibitem[\protect\citeauthoryear{Korol et~al.}{Korol
  et~al.}{2020}]{Korol:2020lpq}
Korol V.,  et~al., 2020, \mn@doi [Astron. Astrophys.]
  {10.1051/0004-6361/202037764}, 638, A153

\bibitem[\protect\citeauthoryear{Kramer \& Champion}{Kramer \&
  Champion}{2013}]{Kramer:2013kea}
Kramer M.,  Champion D.~J.,  2013, \mn@doi [Class. Quant. Grav.]
  {10.1088/0264-9381/30/22/224009}, 30, 224009

\bibitem[\protect\citeauthoryear{Kudoh \& Taruya}{Kudoh \&
  Taruya}{2005}]{Kudoh:2004he}
Kudoh H.,  Taruya A.,  2005, \mn@doi [Phys. Rev. D]
  {10.1103/PhysRevD.71.024025}, 71, 024025

\bibitem[\protect\citeauthoryear{Lasky et~al.}{Lasky
  et~al.}{2016}]{Lasky:2015lej}
Lasky P.~D.,  et~al., 2016, \mn@doi [Phys. Rev. X] {10.1103/PhysRevX.6.011035},
  6, 011035

\bibitem[\protect\citeauthoryear{Lau, Mandel, Vigna-G\'omez, Neijssel,
  Stevenson  \& Sesana}{Lau et~al.}{2020}]{Lau:2019wzw}
Lau M.~Y.,  Mandel I.,  Vigna-G\'omez A.,  Neijssel C.~J.,  Stevenson S.,
  Sesana A.,  2020, \mn@doi [Mon. Not. Roy. Astron. Soc.]
  {10.1093/mnras/staa002}, 492, 3061

\bibitem[\protect\citeauthoryear{Littenberg, Cornish, Lackeos  \&
  Robson}{Littenberg et~al.}{2020}]{Littenberg:2020bxy}
Littenberg T.,  Cornish N.,  Lackeos K.,   Robson T.,  2020, \mn@doi [Phys.
  Rev. D] {10.1103/PhysRevD.101.123021}, 101, 123021

\bibitem[\protect\citeauthoryear{Manchester et~al.}{Manchester
  et~al.}{2013}]{Manchester:2012za}
Manchester R.~N.,  et~al., 2013, \mn@doi [Publ. Astron. Soc. Austral.]
  {10.1017/pasa.2012.017}, 30, 17

\bibitem[\protect\citeauthoryear{Marsh}{Marsh}{2011}]{Marsh_2011}
Marsh T.~R.,  2011, \mn@doi [Classical and Quantum Gravity]
  {10.1088/0264-9381/28/9/094019}, 28, 094019

\bibitem[\protect\citeauthoryear{McMillan}{McMillan}{2011}]{McMillan:2011}
McMillan P.~J.,  2011, \mn@doi [Monthly Notices of the Royal Astronomical
  Society] {10.1111/j.1365-2966.2011.18564.x}, 414, 2446

\bibitem[\protect\citeauthoryear{Meurer et~al.,}{Meurer
  et~al.}{2017}]{Meurer:2017}
Meurer A.,  et~al., 2017, \mn@doi [PeerJ Computer Science]
  {10.7717/peerj-cs.103}, 3, e103

\bibitem[\protect\citeauthoryear{Mingarelli et~al.,}{Mingarelli
  et~al.}{2017}]{Mingarelli:2017fbe}
Mingarelli C. M.~F.,  et~al., 2017, \mn@doi [Nature Astron.]
  {10.1038/s41550-017-0299-6}, 1, 886

\bibitem[\protect\citeauthoryear{Payne, Banagiri, Lasky  \& Thrane}{Payne
  et~al.}{2020}]{Payne:2020pmc}
Payne E.,  Banagiri S.,  Lasky P.,   Thrane E.,  2020, \mn@doi [Phys. Rev. D]
  {10.1103/PhysRevD.102.102004}, 102, 102004

\bibitem[\protect\citeauthoryear{Phinney}{Phinney}{2001}]{Phinney:2001di}
Phinney E.,  2001

\bibitem[\protect\citeauthoryear{Regimbau}{Regimbau}{2011}]{Regimbau:2011rp}
Regimbau T.,  2011, \mn@doi [Res. Astron. Astrophys.]
  {10.1088/1674-4527/11/4/001}, 11, 369

\bibitem[\protect\citeauthoryear{Renzini \& Contaldi}{Renzini \&
  Contaldi}{2018}]{Renzini:2018vkx}
Renzini A.,  Contaldi C.,  2018, \mn@doi [Mon. Not. Roy. Astron. Soc.]
  {10.1093/mnras/sty2546}, 481, 4650

\bibitem[\protect\citeauthoryear{Romano \& Cornish}{Romano \&
  Cornish}{2017}]{Romano:2016dpx}
Romano J.~D.,  Cornish N.~J.,  2017, \mn@doi [Living Rev. Rel.]
  {10.1007/s41114-017-0004-1}, 20, 2

\bibitem[\protect\citeauthoryear{Romano, Taylor, Cornish, Gair, Mingarelli  \&
  van Haasteren}{Romano et~al.}{2015}]{Romano:2015uma}
Romano J.~D.,  Taylor S.~R.,  Cornish N.~J.,  Gair J.,  Mingarelli C. M.~F.,
  van Haasteren R.,  2015, \mn@doi [Phys. Rev. D] {10.1103/PhysRevD.92.042003},
  92, 042003

\bibitem[\protect\citeauthoryear{Schilling}{Schilling}{1997}]{Schilling_1997}
Schilling R.,  1997, \mn@doi [Classical and Quantum Gravity]
  {10.1088/0264-9381/14/6/020}, 14, 1513

\bibitem[\protect\citeauthoryear{Speagle}{Speagle}{2020}]{Speagle_2019}
Speagle J.~S.,  2020, \mn@doi [Monthly Notices of the Royal Astronomical
  Society] {10.1093/mnras/staa278}, 493, 3132

\bibitem[\protect\citeauthoryear{Tamanini \& Danielski}{Tamanini \&
  Danielski}{2019}]{Tamanini_Danielski_2019}
Tamanini N.,  Danielski C.,  2019, \mn@doi [Nature Astronomy]
  {10.1038/s41550-019-0807-y}, 3, 858–866

\bibitem[\protect\citeauthoryear{Taruya}{Taruya}{2006}]{Taruya:2006kqa}
Taruya A.,  2006, \mn@doi [Phys. Rev. D] {10.1103/PhysRevD.74.104022}, 74,
  104022

\bibitem[\protect\citeauthoryear{Taruya \& Kudoh}{Taruya \&
  Kudoh}{2005}]{Taruya:2005yf}
Taruya A.,  Kudoh H.,  2005, \mn@doi [Phys. Rev. D]
  {10.1103/PhysRevD.72.104015}, 72, 104015

\bibitem[\protect\citeauthoryear{Taylor \& Gair}{Taylor \&
  Gair}{2013}]{Taylor:2013esa}
Taylor S.~R.,  Gair J.~R.,  2013, \mn@doi [Phys. Rev. D]
  {10.1103/PhysRevD.88.084001}, 88, 084001

\bibitem[\protect\citeauthoryear{Taylor et~al.}{Taylor
  et~al.}{2015}]{Taylor:2015udp}
Taylor S.~R.,  et~al., 2015, \mn@doi [Phys. Rev. Lett.]
  {10.1103/PhysRevLett.115.041101}, 115, 041101

\bibitem[\protect\citeauthoryear{Taylor, van Haasteren  \& Sesana}{Taylor
  et~al.}{2020}]{Taylor:2020zpk}
Taylor S.~R.,  van Haasteren R.,   Sesana A.,  2020, \mn@doi [Phys. Rev. D]
  {10.1103/PhysRevD.102.084039}, 102, 084039

\bibitem[\protect\citeauthoryear{Thrane, Ballmer, Romano, Mitra, Talukder, Bose
   \& Mandic}{Thrane et~al.}{2009}]{Thrane:2009aa}
Thrane E.,  Ballmer S.,  Romano J.~D.,  Mitra S.,  Talukder D.,  Bose S.,
  Mandic V.,  2009, \mn@doi [Phys. Rev. D] {10.1103/PhysRevD.80.122002}, 80,
  122002

\bibitem[\protect\citeauthoryear{Tinto \& Dhurandhar}{Tinto \&
  Dhurandhar}{2020}]{Tinto:2004wu}
Tinto M.,  Dhurandhar S.~V.,  2020, \mn@doi [Living Rev. Rel.]
  {10.1007/s41114-020-00029-6}, 24, 1

\bibitem[\protect\citeauthoryear{Ungarelli \& Vecchio}{Ungarelli \&
  Vecchio}{2001}]{Ungarelli:2001xu}
Ungarelli C.,  Vecchio A.,  2001, \mn@doi [Phys. Rev. D]
  {10.1103/PhysRevD.64.121501}, 64, 121501

\bibitem[\protect\citeauthoryear{Yang, Mandic, Scarlata  \& Banagiri}{Yang
  et~al.}{2020}]{Yang:2020usq}
Yang K.~Z.,  Mandic V.,  Scarlata C.,   Banagiri S.,  2020, \mn@doi [Mon. Not.
  Roy. Astron. Soc.] {10.1093/mnras/staa3159}, 500, 1666

\bibitem[\protect\citeauthoryear{Zonca, Singer, Lenz, Reinecke, Rosset, Hivon
  \& Gorski}{Zonca et~al.}{2019}]{Zonca2019}
Zonca A.,  Singer L.,  Lenz D.,  Reinecke M.,  Rosset C.,  Hivon E.,   Gorski
  K.,  2019, \mn@doi [Journal of Open Source Software] {10.21105/joss.01298},
  4, 1298

\makeatother
\end{thebibliography}
\bibliographystyle{mnras}

\bsp
\label{lastpage}
\end{document}